\documentclass[11pt]{article}
\usepackage[dvipsnames,table]{xcolor}
\usepackage[preprint]{acl}

\usepackage{times}
\usepackage{latexsym}

\usepackage[T1]{fontenc}

\usepackage[utf8]{inputenc}

\usepackage{microtype}

\usepackage{inconsolata}

\usepackage{caption}
\usepackage{colortbl}
\usepackage{wrapfig} 
\usepackage{amssymb}
\usepackage{multirow}
\usepackage{diagbox}
\usepackage[norelsize, linesnumbered, ruled, lined, boxed, commentsnumbered, noend]{algorithm2e} 
\usepackage{pifont}
\usepackage{booktabs}
\usepackage{boldline}
\usepackage{mdframed}
\usepackage{graphicx}
\usepackage{longtable}
\usepackage{alltt}
\usepackage{enumitem}
\usepackage{amsmath}
\usepackage[most]{tcolorbox}
\usepackage{listings}
\usepackage{epsfig,endnotes}
\title{Planning-Driven Programming: A Large Language Model Programming Workflow}

\author{Chao Lei, Yanchuan Chang, Nir Lipovetzky, Krista A. Ehinger \\
    School of Computing and Information Systems, The University of Melbourne, Australia\\
\texttt{clei1@student.unimelb.edu.au,
}
\\
\texttt{\{yanchuan.chang, nir.lipovetzky, kris.ehinger\}@unimelb.edu.au}}

\begin{document}
\maketitle
\begin{abstract}
The strong performance of large language models (LLMs) raises extensive discussion on their application to code generation. Recent research suggests continuous program refinements through visible tests to improve code generation accuracy in LLMs. However, these methods suffer from LLMs' inefficiency and limited reasoning capacity. In this work, we propose an LLM programming workflow (LPW) designed to improve both initial code generation and subsequent refinements within a structured two-phase workflow. Specifically, the solution generation phase formulates a solution plan, which is then verified through visible tests to  specify the intended natural language solution. Subsequently, the code implementation phase drafts an initial code according to the solution plan and its verification. If the generated code fails the visible tests, the plan verification serves as the intended solution to consistently inform the refinement process for correcting bugs. Compared to state-of-the-art methods across various existing LLMs, LPW significantly improves the Pass@1 accuracy by up to 16.4\% on well-established text-to-code generation benchmarks. LPW also sets new state-of-the-art Pass@1 accuracy, achieving 98.2\% on HumanEval, 84.8\% on MBPP, 59.3\% on LiveCode, 62.6\% on APPS, and 34.7\% on CodeContests, using GPT-4o as the backbone. Our code is publicly available at: https://github.com/you68681/lpw
\end{abstract}

\begin{figure*}[t]
\centering      
\includegraphics[width=1\textwidth]{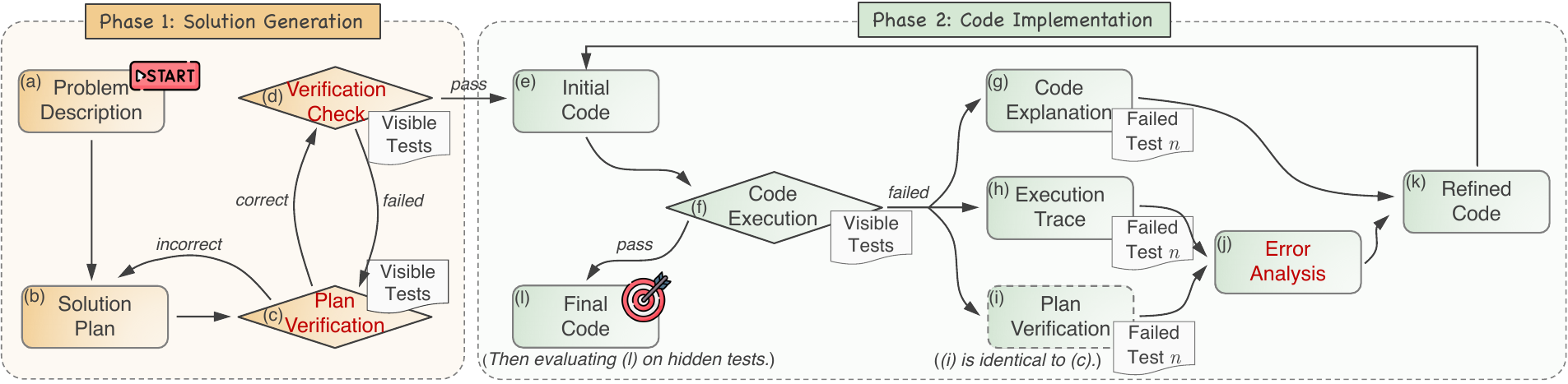}
\vspace{-0.8cm}
\caption{The pipeline of LPW, where the components highlighted in red are exclusive to LPW.} 
\vspace{-0.3cm}
\label{key_component}
\end{figure*}

\section{Introduction}

Code generation, also known as \textit{program synthesis}, studies the automatic construction of a program that satisfies a specified high-level input requirement \citep{gulwani2017program}.  Recently, large language models (LLMs) pre-trained on extensive code-related datasets \citep{brown2020language, meta2024introducing, li2023starcoder, roziere2023code, achiam2023gpt, muennighoff2023octopack} 
have shown success in code-related tasks, such as code generation from natural language descriptions, also named as text-to-code generation~\citep{chen2021evaluating,austin2021program, li2022competition}, code translation \citep{pan2024lost,yang2024exploring}, and code completion \citep{izadi2024language}. However, LLM-based code generation remains challenging due to stringent lexical, grammatical, and semantic constraints \citep{scholak2021picard}.

Code generation substantially benefits from the empirical insights of human programmers. In practice, human programmers develop high-quality code by consistently rectifying errors through the analysis of test case executions \citep{huang2023survey, chen2023teaching}. 
Different studies have refined programs based on execution results and LLM-generated information such as code and error explanation \citep{tang2023explain,shinn2023reflexion,madaan2024self}. 
Recent work further optimizes refinement (debugging) by performing \textit{rubber duck debugging} processes \citep{chen2023teaching} and leveraging control flow graph information to assist LLMs in locating bugs \citep{zhong2024ldb}. However, the absence of precise correction instructions in feedback messages results in numerous refinements that deviate from the intended solution. Additionally, refining programs that significantly diverge from the problem description remains an open challenge \citep{tian2023test}.

To replicate each phase of program development, several studies \citep{lin2024llm,qian2023communicative, dong2023self} have employed LLM instances as customized agents, assigning them diverse roles and facilitating their collaboration. Recent work incorporates  extra visible tests \citep{huang2023agentcoder} and solution plans \citep{islam-etal-2024-mapcoder} generated by designated agents to improve the code refinements in multi-agent collaborations. However, the absence of methodologies for generating reliable visible tests and plans in these studies undermines their credibility, as incorrect visible tests and plans can lead to erroneous codes. Besides, with an increased number of agents, multi-agent collaborations consume significant token resources for communication \citep{huang2023agentcoder}. The detailed related work is discussed in Appendix \ref{Related_Work}.

In this work, we propose LPW, a \textit{large \underline{l}anguage
model \underline{p}rogramming \underline{w}orkflow}, specifically for text-to-code generation, addressing the aforementioned limitations. LPW involves two phases for code generation: the solution generation phase for plan creation and plan verification, and the code implementation phase for initial code development and subsequent refinements. The pipeline of LPW is depicted in Figure \ref{key_component}. LPW leverages various information, including LLM-generated solution plan \citep{jiang2023self} (block (b)), LLM-generated code explanation \citep{chen2023teaching} (block (g)), and runtime information from program execution \citep{zhong2024ldb} (block (h)) to boost the code generation performance, and effectively incorporates them into an end-to-end framework.  In LPW, aside from runtime information, all other messages are autonomously generated by LLMs using few-shot prompting, without the need for annotated corpora or additional training.

A unique feature of LPW is incorporating plan verification (block (c)) as the natural language intended solution for visible tests. LPW initially produces a solution plan that decomposes a complex problem into several tractable sub-problems (intermediate steps) (block (b)). LPW then verifies the solution plan against visible tests to assess its correctness, known as plan verification. For a visible test, the LLM-generated verification includes a text-based step-by-step analysis to derive the output for each intermediate step and the final output, ensuring that the final output is consistent with the visible test result. Subsequently, each inferred intermediate output is reviewed by LLMs (block (d)) to maintain logical consistency and mitigate hallucination throughout the verification. The plan verification encompasses comprehensive conditions and logical specifications for solving visible tests, eliminating potential misunderstandings before code generation.  This is akin to \textit{Test-Driven Development}, where human developers validate the intended solution with test cases \citep{beck2022test}. 

LPW uses the plan and its verification in the initial code development (block (e)) to ensure that the initial code closely aligns with the problem description. Furthermore, LPW incorporates plan verification in the subsequent refinements. By comparing discrepancies between the expected output of each intermediate step, as recorded in the plan verification, against the execution trace on the failed visible test (block (h)), LPW accurately locates bugs, identifies logic flaws in the code implementation, and further generates detailed refinement suggestions, as documented in the error analysis (block (j)). Then, the error analysis when integrated with the code explanation (block (g)) serves as feedback to refine the code, surpassing conventional scalar rewards or deduced error analysis \citep{chen2023teaching, zhong2024ldb, shinn2023reflexion} and thereby improving refinement efficiency and accuracy.

We first evaluate LPW  on four text-to-code generation benchmarks: HumanEval \citep{chen2021evaluating}, MBPP \citep{austin2021program}, and their extended test case variants, HumanEval-ET and MBPP-ET \citep{dong2023codescore}. We conduct experiments on the proprietary LLM GPT-3.5 \citep{achiam2023gpt}, and open-source LLMs, Llama-3 \citep{meta2024introducing} and Phi-3 \citep{abdin2024phi}. The Pass@1 accuracy  \citep{chen2021evaluating} is reported. Compared to the state-of-the-art LLM debugger, LDB \citep{zhong2024ldb}, LPW improves Pass@1 accuracy by 2\% to 6.1\% across all benchmarks with  GPT-3.5 and achieves up to 16.4\% improvement on MBPP with Llama-3.  When evaluated on additional benchmarks using the advanced GPT-4o \citep{openAI2024GPT-4o}, LPW maintains its advantages, and achieves new state-of-the-art performance across evaluated benchmarks. Notably, on the contamination-free benchmark, LiveCode \citep{jain2024livecodebench}, and challenging benchmarks, APPS \citep{hendrycks2021measuring} and CodeContests \citep{li2022competition}, LPW improves Pass@1 accuracy by around 5\%, 10\%, and 5\%, respectively, compared to LDB. We outline our contributions as follows:
\begin{itemize}
    \item We introduce an end-to-end large language model programming workflow, LPW, which significantly improves the code generation accuracy over the state-of-the-art methods.

    \item We derive the intended solution for visible tests, represented as the plan verification. The plan verification clarifies  logic specifications required to solve the visible tests for the given problem, thereby increasing the LLMs' confidence during both the initial program generation and subsequent debugging processes.

    \item We conduct extensive experiments across seven text-to-code generation benchmarks to validate the performance of LPW with various LLM backbones, provide a comprehensive analysis of their performance,  token usage, and failure cases, and highlight the existing challenges.
    
\end{itemize}

\section{Problem Formulation}

We follow the problem formulation for text-to-code generation as outlined in \citet{jiang2023self}, \citet{chen2023teaching}, and \citet{zhong2024ldb}. The text-to-code generation problem is formulated as a triple $\mathcal{P}=\langle Q, T_v, T_h \rangle$, where $Q$ represents the natural language problem specifications, and $T_v$ and $T_h$ are sets of visible and hidden tests, each containing input-output pairs $(t^i, t^o) \in T = T_v \cup T_h$. The goal is to leverage the LLM $\mathcal{M}$ to generate a program function $f$, $\mathcal{M} \rightarrow f$, that maps each input $t^i$ to its output $t^o$ for all pairs in $T$, i.e., $f(t^i)=t^o$, for $(t^i, t^o) \in T$. We note that $T_h$  remains hidden during both solution generation and code implementation phases and only becomes visible if the generated $f$ passes $T_v$.  In LPW, for all components shown in Figure \ref{key_component}, the problem description $Q$ is, by default, concatenated with task-specific prompts to produce the desired response from LLMs.

\begin{figure}
    \centering
    \includegraphics[width=1\linewidth]{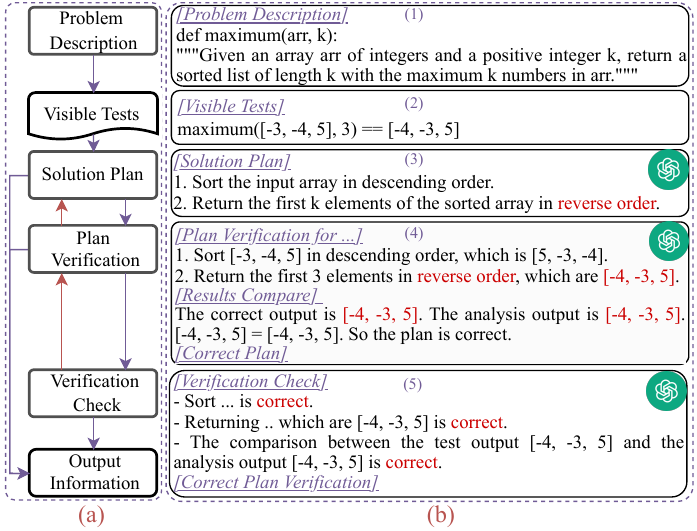}
\vspace{-0.8cm}
\caption{ (a): An illustrated workflow of the solution generation phase in LPW. 
    (b): Example message fragments corresponding to each workflow component for a HumanEval problem (\textit{120th}) with the GPT-3.5 backbone. The detailed messages are available in Section \ref{case_study}.}
    \vspace{-0.2cm}
    \label{workflow_solution_generation}
\end{figure}

\section{Workflow Structure}
In this section, we first detail the two phases of LPW separately and then elaborate on the iterative update strategies used in each phase.

\textbf{Solution Generation}. Figure \ref{workflow_solution_generation} displays the overall workflow of the solution generation phase in LPW (part (a)), with an example programming problem for illustration (part (b)). 
LPW leverages the self-planning approach introduced by \citet{jiang2023self} to abstract and decompose the problem description $Q$ into a strategic and adaptable plan $\Pi$ at the start of the solution generation phase. 

For a problem in HumanEval described by block (1) in Figure \ref{workflow_solution_generation}, its example solution plan is illustrated at block (3).
However, the LLM-generated plan $\Pi$ may occasionally be incorrect, misguiding subsequent program generation. To avoid this, LPW queries the LLM to verify $\Pi$ against all visible tests $T_v$. The LLM-responded plan verification  $\mathcal{A}(\Pi, T_v)$ delivers a step-by-step analysis, including all intermediate results and final derived outputs for all visible tests $T_v$ based on  $\Pi$. 
For each $t_v \in T_v$, its verification $\mathcal{A}(\Pi, \{t_v\})$ compares the derived output $t_v^{o^\prime}$ with the ground-truth output $t_v^o$ to assess the correctness of $\Pi$, as outlined at block 4. If $\Pi$ is successfully verified on all visible tests, where in $\mathcal{A}(\Pi, T_v)$, $t_v^{o^\prime}=t_v^o, \forall t_v\in T_v$, then the plan verification $\mathcal{A}(\Pi, T_v)$ is reviewed by the LLM again to ensure the accuracy of all intermediate results, since  each intermediate step result is used in locating bugs and providing refinement suggestions when compared with the code runtime information on the failed visible test. If all intermediate outputs in $\mathcal{A}(\Pi, T_v)$ are validated as correct by the LLM as shown at block 5, $\mathcal{A}(\Pi, T_v)$ is treated as the intended solution for $T_v$. The plan $\Pi$ and its verification $\mathcal{A}(\Pi, T_v)$ serve as the output of the solution generation phase, guiding code development and refinements in the code implementation phase.

\begin{figure}
    \centering
    \includegraphics[width=1\linewidth]{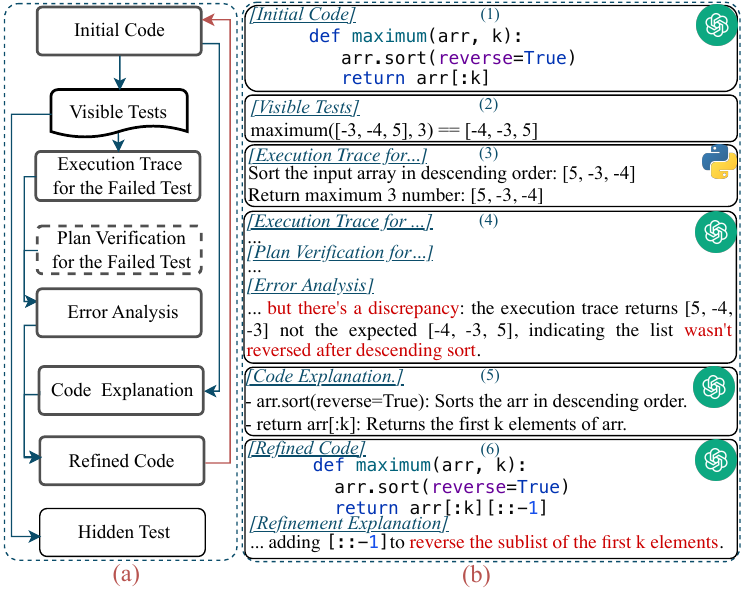}
    \vspace{-0.8cm}
    \caption{(a): An illustrated workflow of the code implementation phase in LPW. (b): Example message fragments extending from Figure \ref{workflow_solution_generation} and corresponding to each workflow component. 
    See Section \ref{case_study} for detailes.
    } \label{workflow_code_implemtation}
        \vspace{-0.5cm}
\end{figure}

\textbf{Code Implementation}. Figure \ref{workflow_code_implemtation} shows the overall workflow of the code implementation phase in LPW (part (a)), using the same problem from Figure \ref{workflow_solution_generation} as an illustration (part (b)). 
LPW develops an initial program $f$ by prompting the LLM with the problem description $Q$ (block (1) in Figure \ref{workflow_solution_generation}), along with plan $\Pi$ and its verification $\mathcal{A}(\Pi, T_v)$ from the solution generation phase. Subsequently, LPW queries the LLM to add \textit{print statements} for each line in $f$, resulting in $f_p$, and then executes $f_p$ on visible tests $T_v$. If $f_p$ successfully solves $T_v$, LPW validates it on the hidden tests $T_h$ to report Pass@1 accuracy. 
Otherwise,  LPW collects the runtime information on the first failed visible test $\Bar{t_v}$, indicating that the implementation in $f$ deviates from the specifications in $\mathcal{A}(\Pi, \{\Bar{t_v}\})$. Blocks 1-3 in part (b) of Figure \ref{workflow_code_implemtation} depict an initial program $f$ (block (1)) that fails on a visible test $\Bar{t_v}$ (block (2)) and its execution trace (block (3)) on $\Bar{t_v}$ after adding print statements. We omit $f_p$ from Figure \ref{workflow_code_implemtation} to keep the discussion concise. LPW instructs the LLM to conduct an error analysis by identifying inconsistencies between the intermediate outputs recorded in the execution trace of $\Bar{t_v}$ and the expected intermediate outputs documented in the verification $\mathcal{A}(\Pi,\{\Bar{t_v}\})$, analyzing causes, and offering refinement suggestions (block (4)). Subsequently, the error analysis and code explanation for $f$ generated by the LLM (block (5)) are concatenated as the prompt to generate the refined program $f^\prime$ (block (6)). The code explanation helps the LLM align the text-based error analysis with the code implementation. LPW replaces $f$ with the refined program $f^\prime$ and revalidates the updated $f$ against the visible tests $T_v$ to assess the need for further refinements.

\textbf{Iterative Updates}. LPW includes two update steps in the solution generation phase to enable \textit{self-correction} as indicated by the red arrows in  Figure \ref{workflow_solution_generation}: 1) when the plan verification inferred final output differs from the ground-truth output for a visible test, where $t_v^{o^\prime} \neq t_v^o, \exists t_v \in T_v$ in $\mathcal{A}(\Pi, T_v)$, a revised solution plan $\Pi^\prime$ is included in the LLM response to substitute the original plan; 2) when the LLM detects any incorrect intermediate values in $\mathcal{A}(\Pi, T_v)$ e.g., 
contextual inconsistencies, mathematical miscalculations, or logical flaws, LPW prompts the LLM to regenerate the plan verification. These update methods ensure that the solution plan $\Pi$ and its verification $\mathcal{A}(\Pi, T_v)$ maintain the necessary precision, as well-formed $\Pi$ and $\mathcal{A}(\Pi, T_v)$ are essential for accurate code generation \citep{jiang2023self}. In the code implementation phase, the code refinement process acts as an update mechanism, replacing the program $f$ with the refined program  $f^\prime$ when $f$ fails the visible test $T_v$ as highlighted by the red arrow in Figure \ref{workflow_code_implemtation}. Overall, for a problem $\mathcal{P}$, LPW iteratively revises the generated plan $\Pi$ and its verification $\mathcal{A}(\Pi, T_v)$, in the solution generation phase, until $\mathcal{A}(\Pi, T_v)$ infers correct outputs for all visible tests $T_v$ and no error intermediate outputs are present in $\mathcal{A}(\Pi, T_v)$. Otherwise, LPW reports a failure for $\mathcal{P}$ when reaching the maximum iterations. Similarly, in the code implementation phase, LPW iteratively refines the  generated program $f$ if bugs exist. This process continues until a refined $f$ successfully solves $T_v$, followed by Pass@1 accuracy calculation on hidden tests $T_h$, or LPW reports a failure for $\mathcal{P}$ upon reaching the maximum iteration limit.

\begin{table*}[t]
\centering
\small
\setlength{\tabcolsep}{8.5pt}
\renewcommand{\arraystretch}{1}
\begin{tabular}{llcccccccc}
\toprule
&            & \multicolumn{2}{c}{HumanEval}                   & \multicolumn{2}{c}{HumanEval-ET}                      & \multicolumn{2}{c}{MBPP}                              & \multicolumn{2}{c}{MBPP-ET}                           \\
                          \cmidrule(lr){3-4}  \cmidrule(lr){5-6}  \cmidrule(lr){7-8} \cmidrule(lr){9-10}
                          &            & \multicolumn{1}{c}{Acc $\uparrow$}   & \multicolumn{1}{c}{$\Delta$ $\uparrow$}  & \multicolumn{1}{c}{Acc $\uparrow$}         & \multicolumn{1}{c}{$\Delta$ $\uparrow$}  & \multicolumn{1}{c}{Acc $\uparrow$}       & \multicolumn{1}{c}{$\Delta$ $\uparrow$}  & \multicolumn{1}{c}{Acc $\uparrow$}         & \multicolumn{1}{c}{$\Delta$ $\uparrow$} \\  \hline\hline
                          & Baseline   & 74.4 {\scriptsize $\pm 0.8$}                 & \multicolumn{1}{c}{--} & 66.5  {\scriptsize $\pm 1.3$}                                & \multicolumn{1}{c}{--} & 67.4  {\scriptsize $\pm 0.5$}                                & \multicolumn{1}{c}{--} & 52.8  {\scriptsize $\pm 0.3$}                                & \multicolumn{1}{c}{--} \\
                          & SP         & 77.4   {\scriptsize $\pm 0.8$}                         & 3.0                 & 69.5 {\scriptsize $\pm 0.8$}                                 & 3.0                 & 69.2  {\scriptsize $\pm 0.4$}                                & 1.8                 & 52.4    {\scriptsize $\pm 0.2$}                              & \textcolor{ForestGreen}{-0.4}                \\
                          & MapCoder         & 77.4   {\scriptsize $\pm 1.0$}                         & 3.0                 & 66.5 {\scriptsize $\pm 1.0$}                                 & 0.0                 & 72.0  {\scriptsize $\pm 0.5$}                                & 4.6                 & 56.6    {\scriptsize $\pm 0.5$}                              & 3.8                \\
                          
                          & SD         & 81.1  {\scriptsize $\pm 1.0$}                          & 6.7                 & 72.0 {\scriptsize $\pm 1.0$}                                  & 5.5                 & 71.2 {\scriptsize $\pm 0.3$}                                 & 3.8                 & 56.0 {\scriptsize $\pm 0.1$}                                 & 3.2                 \\
                          & LDB               & 82.9 {\scriptsize $\pm 1.0$}                    & 8.5        & 72.6   {\scriptsize $\pm 1.0$}                                  & 6.1                 & 72.4   {\scriptsize $\pm 0.3$}                                & 5.0                 & 55.6   {\scriptsize $\pm 0.2$}                                & 2.8                 \\
                          \multirow{-5.75}{*}{GPT-3.5} & \cellcolor[HTML]{EEEEEE}\textbf{LPW (ours)}  & \cellcolor[HTML]{EEEEEE}\textbf{\textcolor{red}{89.0  {\scriptsize $\pm 0.8$}}}                  & \cellcolor[HTML]{EEEEEE}\textbf{\textcolor{red}{14.6}}                & \cellcolor[HTML]{EEEEEE}\textbf{\textcolor{red}{77.4 \scriptsize $\pm 0.8$}}                        & \cellcolor[HTML]{EEEEEE}\textbf{\textcolor{red}{10.9}}                & \cellcolor[HTML]{EEEEEE}\textbf{\textcolor{red}{76.0 \scriptsize $\pm 0.2$}}                        & \cellcolor[HTML]{EEEEEE}\textbf{\textcolor{red}{8.6}}                 & \cellcolor[HTML]{EEEEEE}\textbf{\textcolor{red}{57.6 \scriptsize $\pm 0.1$}}                        & \cellcolor[HTML]{EEEEEE}\textbf{\textcolor{red}{4.8}}                 \\ \hline
                          & Baseline   & 73.2  \scriptsize $\pm 1.3$                & \multicolumn{1}{c}{--} & 61.0 \scriptsize $\pm 1.0$                       & \multicolumn{1}{c}{--} & 44.0 \scriptsize $\pm 1.2$                       & \multicolumn{1}{c}{--} & 35.4 \scriptsize $\pm 1.0$                       & \multicolumn{1}{c}{--} \\
                          & SP         & 78.0 \scriptsize $\pm 2.0$                  & 4.8                 & 65.2 \scriptsize $\pm 1.0$                        & 4.2                 & 48.6   \scriptsize $\pm 1.4$                      & 4.6                 & 38.4   \scriptsize $\pm 1.4$                      & 3.0                 \\
                          & MapCoder         & 83.5   {\scriptsize $\pm 1.3$}                         & 10.3                 & 73.8 {\scriptsize $\pm 0.8$}                                 & 12.8                 &71.4 {\scriptsize $\pm 1.0$}                                & 27.4                 & 55.6    {\scriptsize $\pm 1.0$}                              & 20.2                \\
                          & SD         & 81.7 \scriptsize $\pm 1.3$                  & 8.5                 & 68.3   \scriptsize $\pm 0.8$                      & 7.3                 & 63.6  \scriptsize $\pm 1.2$                       & 19.6                & 50.0   \scriptsize $\pm 1.3$                      & 14.6                \\
                          & LDB        & 84.1   \scriptsize $\pm 1.7$                & 10.9                & 72.0  \scriptsize $\pm 0.8$                        & 11.0                & 57.2 \scriptsize $\pm 1.6$                         & 13.2                & 44.8  \scriptsize $\pm 1.4$                       & 9.4                 \\
                          \multirow{-5.75}{*}{Llama-3}  &\cellcolor[HTML]{EEEEEE}\textbf{LPW (ours)}  & \cellcolor[HTML]{EEEEEE}\textbf{\textcolor{red}{88.4 \scriptsize $\pm 1.6$ }}                  & \cellcolor[HTML]{EEEEEE}\textbf{\textcolor{red}{15.2}}                & \cellcolor[HTML]{EEEEEE}\textbf{\textcolor{red}{76.2 \scriptsize $\pm 1.3$ }}                        & \cellcolor[HTML]{EEEEEE}\textbf{\textcolor{red}{15.2}}                & \cellcolor[HTML]{EEEEEE}\textbf{\textcolor{red}{73.6 \scriptsize $\pm 1.3$ }}                        & \cellcolor[HTML]{EEEEEE}\textbf{\textcolor{red}{29.6}}                & \cellcolor[HTML]{EEEEEE}\textbf{\textcolor{red}{56.4 \scriptsize $\pm 1.2$ }}                        & \cellcolor[HTML]{EEEEEE}\textbf{\textcolor{red}{21.0}}                \\\hline
                          & Baseline   & 36.0 \scriptsize $\pm 1.0$                  & \multicolumn{1}{c}{--}  & 32.3  \scriptsize $\pm 1.0$                       & \multicolumn{1}{c}{--}  & 39.0 \scriptsize $\pm 1.3$                        & \multicolumn{1}{c}{--}  & 33.2  \scriptsize $\pm 1.4$                       & \multicolumn{1}{c}{--}  \\
                          & SP         & 40.8    \scriptsize $\pm 1.4$               & 4.8                 & 34.8  \scriptsize $\pm 0.9$                       & 2.5                 & 46.4   \scriptsize $\pm 1.4$                      & 7.4                 & 37.6  \scriptsize $\pm 1.4$                       & 4.4                 \\
                          & MapCoder         & --                          & --                 & --                               & --                 & --                              & --                & --                           & --                \\
                          & SD         & 51.2   \scriptsize $\pm 1.2$                & 15.2                & 45.7  \scriptsize $\pm 1.0$                       & 13.4                & \multicolumn{1}{c}{45.8 \scriptsize $\pm 1.2$}          & \multicolumn{1}{c}{6.8}  & \multicolumn{1}{c}{36.6 \scriptsize $\pm 1.2$}          & \multicolumn{1}{c}{3.4}  \\
                          & LDB        & 65.9  \scriptsize $\pm 1.6$                 & 29.9                & 54.9   \scriptsize $\pm 0.9$                      & 22.6                & 52.4  \scriptsize $\pm 1.6$ & 13.4                & \multicolumn{1}{c}{42.8 \scriptsize $\pm 1.4$ }          & \multicolumn{1}{c}{9.6}  \\
                          \multirow{-5.75}{*}{Phi-3}  & \cellcolor[HTML]{EEEEEE}\textbf{LPW (ours)}  & \cellcolor[HTML]{EEEEEE}\textbf{\textcolor{red}{76.8 \scriptsize $\pm 1.3$ }}                  & \cellcolor[HTML]{EEEEEE}\textbf{\textcolor{red}{40.8}}                & \cellcolor[HTML]{EEEEEE}\textbf{\textcolor{red}{62.8 \scriptsize $\pm 1.3$ }}                        & \cellcolor[HTML]{EEEEEE}\textbf{\textcolor{red}{30.5}}                & \cellcolor[HTML]{EEEEEE}\textbf{\textcolor{red}{64.0 \scriptsize $\pm 1.2$ }}                        &\cellcolor[HTML]{EEEEEE}\textbf{\textcolor{red}{25.0}}                &\cellcolor[HTML]{EEEEEE}\textbf{\textcolor{red}{48.4 \scriptsize $\pm 1.2$}}                        &\cellcolor[HTML]{EEEEEE}\textbf{\textcolor{red}{15.2}}                \\
\hlineB{2}
\end{tabular}
\vspace{-0.3cm}
\caption{Comparisons of Baseline, SP, MapCoder, SD, LDB, and LPW in terms of Pass@1 accuracy (Acc) and improvement ($\Delta$) with respect to Baseline across benchmarks HumanEval, HumanEval-ET, MBPP, and MBPP-ET with LLMs GPT-3.5, Llama-3, and Phi-3. Acc and $\Delta$ are measured in percentages. Best results are highlighted in red. The standard deviation ($\pm$) is calculated based on three runs and applies to other experiments when reported. }
\vspace{-0.3cm}
\label{table:detail_information}
\end{table*}

\section{Experiments}\label{sec:exp}
\textbf{Benchmarks}. 
We first evaluate LPW on the well-established text-to-code benchmarks HumanEval, MBPP, HumanEval-ET, and MBPP-ET, where the given context outlines the intended functionality of the program to be synthesized. HumanEval-ET and MBPP-ET introduce approximately 100 additional hidden tests, covering numerous edge cases, for each problem in HumanEval and MBPP, thus being regarded as more reliable benchmarks ~\citep{dong2023codescore}. In HumanEval and HumanEval-ET, we treat the test cases described in the task description as visible tests, typically 2-5 per task. For MBPP, we consider its test set that contains 500 problems with 3 hidden tests per problem.
We set the first hidden test as the visible test and treat the other two as hidden, consistent with studies \citep{chen2023teaching, zhong2024ldb, ni2023lever,shi2022natural}. MBPP-ET uses the same set of problems and visible tests for each problem as MBPP.

\textbf{Experimental Setup}. We compare LPW with the representative code generation approaches \textit{Self-Planning} (SP)~\citep{jiang2023self}, \textit{MapCoder} ~\citep{islam-etal-2024-mapcoder}, \textit{Self-Debugging (+Expl)} (SD) ~\citep{chen2023teaching}, and \textit{Large Language Model Debugger} (LDB)~\citep{zhong2024ldb}. SP relies solely on the LLM-generated solution plan to produce the program solution in a single effort without refinements. MapCoder, a multi-agent collaborative approach, generates multiple unverified plans and traverses them to produce and refine code based on the current plan. SD uses a \textit{rubber duck debugging} approach in LLMs, where LLMs are prompted to provide explanations of generated programs as feedback for debugging. LDB, a state-of-the-art LLM debugger, segments generated programs into blocks based on the control flow graph, which facilitates bug detection and the refinement of each program block using runtime information in LLMs. A detailed comparison of different methods is summarized in Appendix Table \ref{features_stratgies}.

We generate a seed program for each problem with the same prompts and parameters introduced by \citet{chen2023teaching}  for  SD and LDB and label the performance of seed programs as Baseline. We note that SD and LDB only perform refinements on the seed program that fails the visible tests. We experiment with various LLMs with different parameter sizes, including GPT-3.5 (turbo-0125, $\geq$175B), Llama-3 (70B-Instruct), and Phi-3 (14B-Instruct) to evaluate performance and demonstrate that LPW is model-independent. The Pass@1 accuracy is reported. We apply 2-shot prompting in LPW, with a maximum of 12 iterations for both the solution generation and code implementation phases. Similarly, we set the maximum debugging iterations to 12 for SD and LDB. MapCoder generates 3 plans, each with up to 4 refinement iterations. All following experiments adhere to these iteration settings. An empirical discussion on the number of iterations for LPW is available in Appendix \ref{Parameter_study}.

\textbf{Results on various LLMs}. Table \ref{table:detail_information} presents the Pass@1 accuracy for evaluated approaches, along with their respective improvements over Baseline. LPW outperforms all competing methods across all benchmarks and with various LLM backbones. Compared to LDB, LPW improves Pass@1 accuracy by 6.1\%, 4.8\% 3.6\%, and 2\%, on HumanEval, HumanEval-ET, MBPP, and MBPP-ET, respectively, with GPT-3.5 and achieves up to 16.4\% improvement on MBPP with Llama-3. These results showcase the effectiveness of the proposed workflow and demonstrate the model-independent benefits of LPW. MapCoder fails on Phi-3 as it requires strict XML-formatted responses, which pose a challenge for Phi-3. The failure analysis for LPW with GPT-3.5 is available in Appendix \ref{error_GPT-3.5}. 

\begin{table*}[t]
\centering
\small
\setlength{\tabcolsep}{5pt}
\renewcommand{\arraystretch}{1}
\begin{tabular}{llccccccc}
\hlineB{2}
              & & {HumanEval}                                    & {HumanEval-ET} & {MBPP} &{MBPP-ET}  &{LiveCode}                                        &{APPS}                              &{CodeContests}                           \\ 
                          \hline\hline
              & Baseline  & 91.5 \scriptsize $\pm 0.3$  & 81.7 \scriptsize $\pm 0.3$                                       & 78.4  \scriptsize $\pm 0.4$  & 62.6 \scriptsize $\pm 0.2$  & 45.7  \scriptsize $\pm 0.6$            & 41.7\scriptsize$\pm 0.9$   &28.0 \scriptsize$\pm 0.5$    \\ 
              \multirow{-1}{*}{GPT-4o}  & LDB       & 92.1  \scriptsize$\pm 0.0$     & 81.7 \scriptsize $\pm 0.0$                            & 82.4   \scriptsize$\pm 0.3$   & 65.4 \scriptsize $\pm 0.0$ &   54.3   \scriptsize$\pm 0.3$   & 53.2 \scriptsize$\pm 0.3$&     29.3 \scriptsize$\pm 0.3$       \\  & \cellcolor[HTML]{EEEEEE}\textbf{LPW (ours)}  & \cellcolor[HTML]{EEEEEE}\textbf{\textcolor{red}{98.2 \scriptsize$\pm 0.3$}}  & \cellcolor[HTML]{EEEEEE}\textbf{\textcolor{red}{84.8 \scriptsize$\pm 0.3$}}                              & \cellcolor[HTML]{EEEEEE}\textbf{\textcolor{red}{84.8 \scriptsize$\pm 0.2$}}  & \cellcolor[HTML]{EEEEEE}\textbf{\textcolor{red}{65.8 \scriptsize$\pm 0.1$}}  & \cellcolor[HTML]{EEEEEE}\textbf{\textcolor{red}{59.3 \scriptsize$\pm 0.6$}}                & \cellcolor[HTML]{EEEEEE}\textbf{\textcolor{red}{62.6 \scriptsize$\pm 0.3$}} &    \cellcolor[HTML]{EEEEEE}\textbf{\textcolor{red}{34.7 \scriptsize$\pm 0.3$}}           \\
\hlineB{2}
\end{tabular}
\vspace{-0.3cm}
\caption{Pass@1 accuracy, in percentages, for Baseline, LDB, and LPW on HumanEval, HumanEval-ET, MBPP, MBPP-ET, LiveCode, APPS and CodeContests benchmarks when using GPT-4o (2024-05-13) as the backbone.} 
\vspace{-0.6cm}
\label{table:gpt4o}
\end{table*}

\textbf{Results on Advanced LLM}. To further demonstrate the effectiveness of LPW, we evaluate its performance on the same benchmarks presented in Table \ref{table:detail_information}, along with the contamination-free benchmark, LiveCode, and two competitive benchmarks, APPS and CodeContests, using the advanced LLM GPT-4o as the backbone. LDB is compared due to its second-highest performance in Table \ref{table:detail_information} using GPT-3.5. We sample 140 problems from LiveCode, published between November 2023 and September 2024, postdating GPT-4o's cutoff date. For APPS and CodeContests, we use subsets of 139 and 150 problems, respectively. The experiment results are shown in Table \ref{table:gpt4o}. Similarly, the performance of the seed programs for LDB is referred to as Baseline. LPW outperforms Baseline and LDB across all benchmarks, and establishes new state-of-the-art Pass@1 accuracy, notably achieving 98.2\% on HumanEval. GPT-4o exhibits reduced performance on LiveCode, while LPW reliably outperforms LDB by 5\% accuracy. For APPS and CodeContests, LPW surpasses LDB by  around 10\% and 5\% accuracy, highlighting the advantages of LPW in tackling challenging benchmarks. Incorporating the plan verification allows LPW to clarify issues before code generation and efficiently correct bugs overlooked by LLMs. In contrast, LDB shows a negligible improvement of only 0.6\% and 1.3\% compared to Baseline on HumanEval and CodeContests, underscoring the limitations of debugging with coarse feedback.  See Appendix~\ref{error_GPT-4o} for tasks that LPW fails to address with GPT-4o.

\begin{table}
\setlength{\tabcolsep}{3.2pt}
\renewcommand{\arraystretch}{1}
\small
\begin{tabular}{llccc}
\hlineB{2}
          &    & MBPP-ET $\uparrow$ & MBPP-ET-3 $\uparrow$ &    $\Delta$ $\uparrow$   \\\hline \hline
&MapCoder    & 56.6 \scriptsize$\pm 0.5$ & 60.6 \scriptsize$\pm 0.2$    & 4.0 \\
&SD    & 56.0 \scriptsize$\pm 0.1$ & 59.2 \scriptsize$\pm 0.3$    & 3.2 \\
 \multirow{-1.75}{*}{GPT-3.5}  &LDB          & 55.6 \scriptsize$\pm 0.2$  & 57.6 \scriptsize$\pm 0.2$   & 2.0 \\
&\cellcolor[HTML]{EEEEEE}\textbf{LPW (ours)}      & \cellcolor[HTML]{EEEEEE}\textbf{\textcolor{red}{57.6 \scriptsize$\pm 0.1$}}  & \cellcolor[HTML]{EEEEEE}\textbf{\textcolor{red}{62.0 \scriptsize$\pm 0.2$}}    & \cellcolor[HTML]{EEEEEE}\textbf{\textcolor{red}{4.4}} \\
\hlineB{2}
\end{tabular}
\vspace{-0.3cm}
\caption{The impact on Pass@1 accuracy with additional visible tests using the GPT-3.5 backbone. MBPP-ET-3 includes two more visible tests per problem than MBPP-ET. $\Delta$ represents the accuracy improvement on MBPP-ET-3 over MBPP-ET. Pass@1 accuracy and $\Delta$ are measured as percentages.}
\vspace{-0.2cm}

\label{table:learn_from_test}
\end{table}

\begin{figure}[t]
    \begin{center}    
        \includegraphics[width=0.49\textwidth]{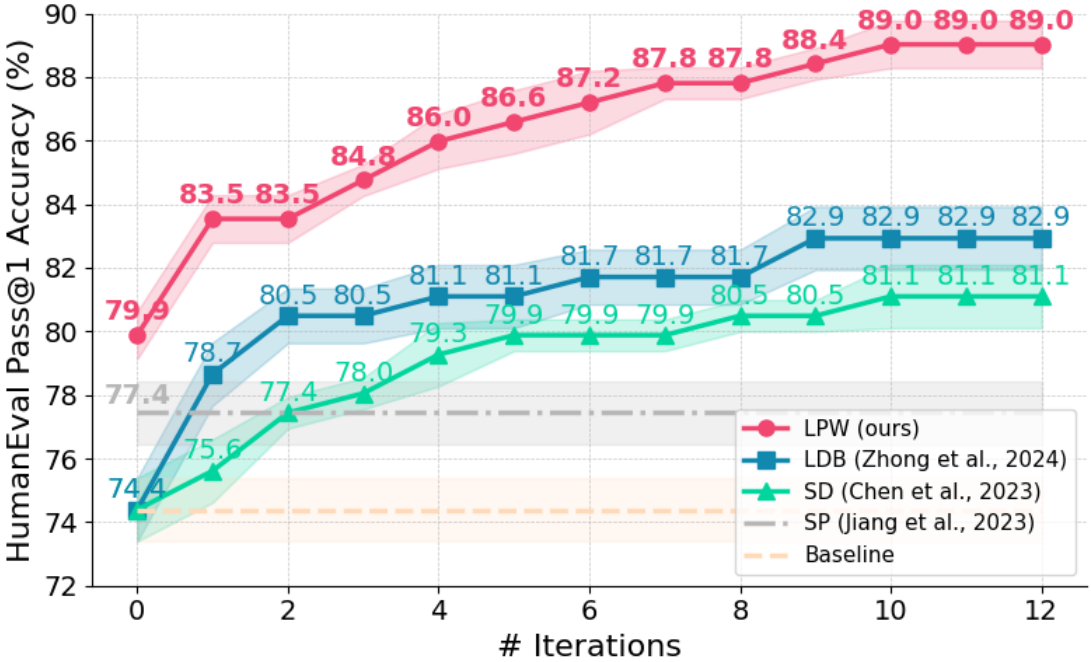}   
    \end{center}

\vspace{-0.5cm}
\caption{The impact on Pass@1 accuracy with the increased number of code implementation iterations/debugging iterations on the HumanEval benchmark when leveraging GPT-3.5 as the LLM backbone. The shaded area represents the standard deviation.}
    \vspace{-0.4cm}
\label{graph_com}
\end{figure}

\textbf{Learning from Test}. We further investigate the impact of the number of visible tests on MapCoder, SD, LDB, and LPW that use visible tests to refine code. We propose a variant of MBPP-ET, denoted as MBPP-ET-3. In MBPP-ET-3, each problem’s visible tests are the three hidden tests from MBPP, while the hidden tests are the extended test cases introduced in MBPP-ET. In other words, each problem in MBPP-ET-3 contains two more visible tests than in MBPP-ET. Results in Table~\ref{table:learn_from_test} show that LPW achieves the highest Pass@1 accuracy of 62.0\% on MBPP-ET-3 and the largest improvement of 4.4\%  over MBPP-ET. LPW produces the step-by-step solution for each visible test to clarify code development logic and inform subsequent refinements, demonstrating superior efficiency in utilizing visible tests to improve performance among evaluated methods.

\textbf{Performance Analysis}. Figure \ref{graph_com} evaluates the Pass@1 accuracy of LPW when considering different numbers of code implementation iterations on the HumanEval benchmark with GPT-3.5. For SD and LDB, we allocate the same number of debugging iterations. We omit MapCoder due to its distinct refinement strategy. We note that all evaluated approaches start from iteration 0, representing the Pass@1 accuracy before debugging. In Figure \ref{graph_com}, Baseline and SP are plotted as straight lines with 74.4\% and 77.4\% accuracy, respectively, due to no debugging involved. Baseline and SP serve as the control group to illustrate when debugging methods surpass no-debugging methods. SD and LDB refine incorrect programs in Baseline, surpassing SP after two iterations. LPW starts debugging from an initial 79.9\% accuracy, higher than the 77.4\% for SP, underscoring the importance of plan verification in initial code generation. LPW surpasses the best performance of SD and LDB after only one iteration, demonstrating its efficient code refinement strategy. LPW gradually refines the code and reaches the highest accuracy by the 10th iteration.

\begin{table}
\setlength{\tabcolsep}{5.2pt}
\renewcommand{\arraystretch}{1}
\small
\begin{tabular}{llcccc}
\toprule
      & & \multicolumn{2}{c}{HumanEval}                       & \multicolumn{2}{c}{MBPP}                    \\
       \cmidrule(lr){3-4} \cmidrule(lr){5-6}
      & & \multicolumn{1}{c}{Acc }         & \multicolumn{1}{c}{$\Delta$} & \multicolumn{1}{c}{Acc} & \multicolumn{1}{c}{$\Delta$} \\\hline\hline
&LPW    & 89.0 \scriptsize $\pm 0.8$                   & --                    & 76.0 \scriptsize $\pm 0.2$                & --                    \\
 &LPW-V  &   86.0   \scriptsize $\pm 0.5$                      & -3.0                & 73.2 \scriptsize $\pm 0.2$                   & -2.8                     \\
 \multirow{-1.75}{*}{GPT-3.5} &LPW-S  &   86.0 \scriptsize $\pm 1.0$                        & -3.0                & 73.0   \scriptsize $\pm 0.3$                 & -3.0               \\
&LPW-C  & 79.9  \scriptsize $\pm 0.8$                       & -9.1                  &   72.2 \scriptsize $\pm 0.5$               &   -3.8               \\
\hlineB{2}
\end{tabular}
\vspace{-0.3cm}
\caption{Pass@1 accuracy (Acc) for different variants of LPW with GPT-3.5. $\Delta$ denotes the decrease against LPW. Acc and $\Delta$ are measured in percentages. See Appendix \ref{Additional_Ablation_Study} for additional ablation study.}
\vspace{-0.4cm}
\label{table:ablation}

\end{table}

\begin{figure}[t]
    \centering
    \includegraphics[width=1\linewidth]{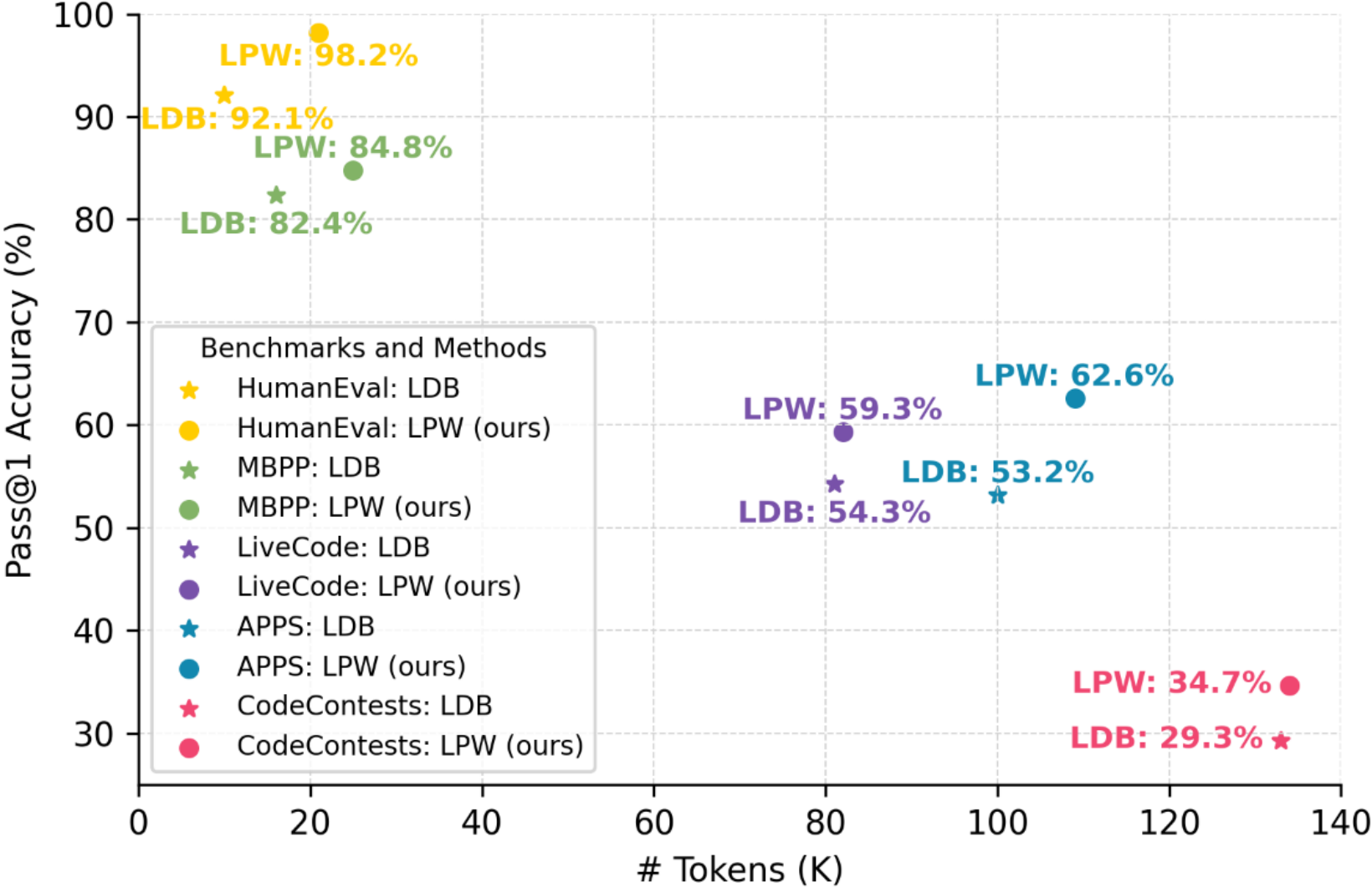}
    \vspace{-0.8cm}
    \caption{Pass@1 accuracy vs. average token cost per program for LDB and LPW on different benchmarks using GPT-4o as the LLM backbone. K is $10^3$. The standard deviation is too small to be visible.
    }
        \vspace{-0.3cm}
    \label{token_cost_whole}
\end{figure}

\begin{figure}[t]
    \centering
        \includegraphics[width=1\linewidth]{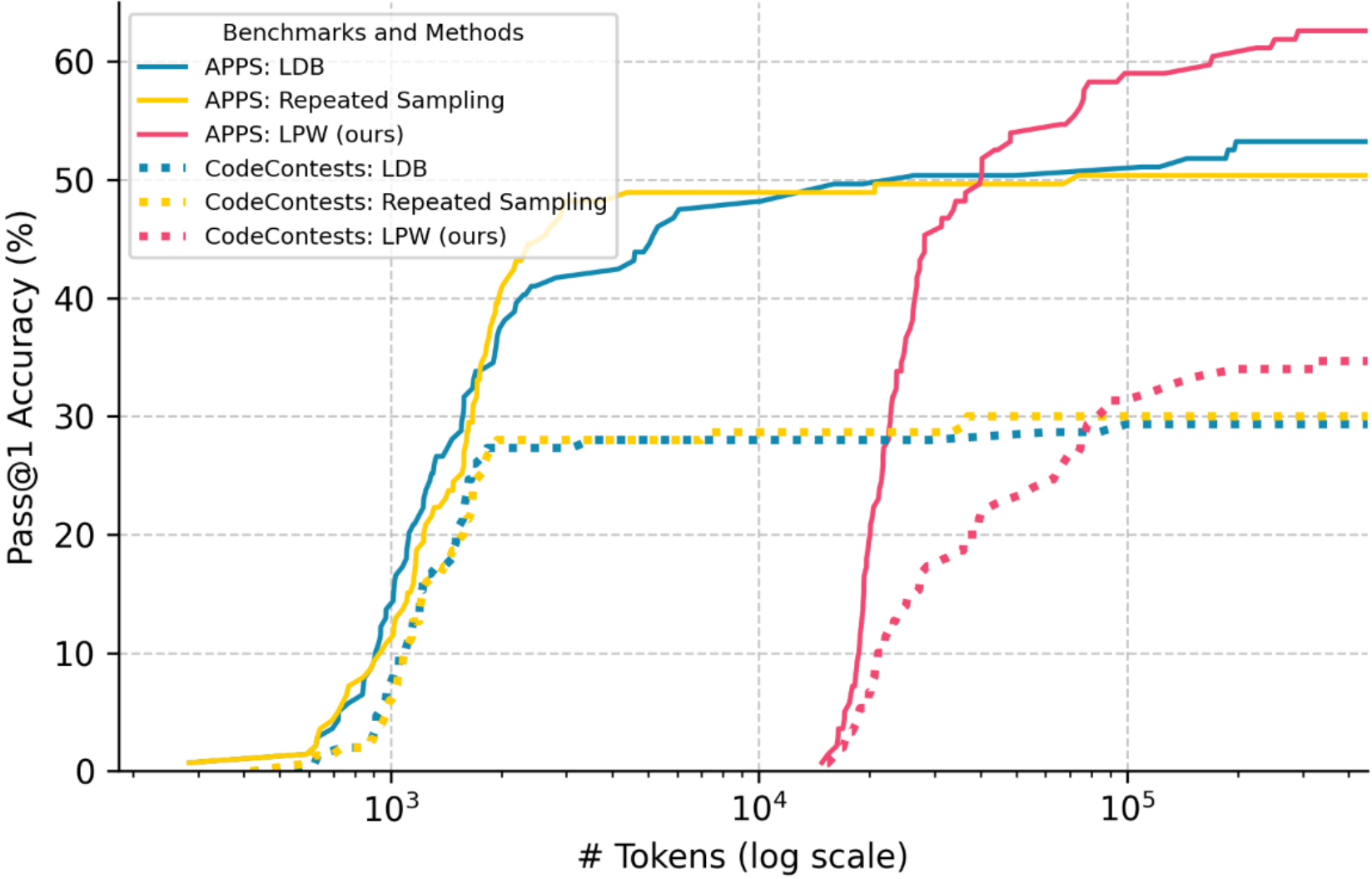}
        \vspace{-0.8cm}
    \caption{Pass@1 accuracy as a function of token consumption for LDB, Repeated Sampling, and LPW  on the APPS and CodeContests benchmarks with GPT-4o. The standard deviation is omitted for better illustration. The same illustration for the LiveCode benchmark is available in Appendix Figure \ref{LiveCode_Token}.}
       \vspace{-0.3cm}
    \label{token_performance}
\end{figure}

\textbf{Ablation Study}. Table \ref{table:ablation} shows the Pass@1 accuracy of different variants of LPW on the HumanEval and MBPP benchmarks with GPT-3.5. The suffix -V denotes the exclusion of plan verification in both solution generation and code implementation phases; -S stands for the LPW variant that excludes the solution generation phase; while -C represents the removal of the code implementation phase, specifically omitting code refinements. For each problem, LPW-V generates the initial program based on the unverified plan and repairs the program using only code explanation and runtime information. LPW-S refines the seed program from Baseline that fails visible tests, leveraging only code explanation and runtime information. LPW-C generates the program solution based on the plan and its verification without refinements. 

The performance decline of LPW-V highlights the significance of plan verification, which serves as the intended solution for visible tests, improving the performance of LLMs in both initial code generation and subsequent refinements. LPW-V considers the unverified plan when drafting initial programs. However, the effect of the unverified plan is limited, as LPW-V shows only slight improvement on MBPP compared to LPW-S, which excludes both the plan and plan verification. This aligns with the results in Table \ref{table:detail_information}, where Self-Planning shows minimal improvement compared to Baseline. The results of LPW-S and LPW-C show that removing either phase in LPW decreases its performance, indicating that both solution generation and code implementation phases are crucial for optimal performance. See Appendix \ref{output_gpt-3.5} and \ref{Consistency}  for a discussion on the accuracy of  LLM-generated plans, plan verifications and refined programs in LPW.

\begin{figure*}[t]
    \begin{center}     
    \includegraphics[width=1\textwidth]{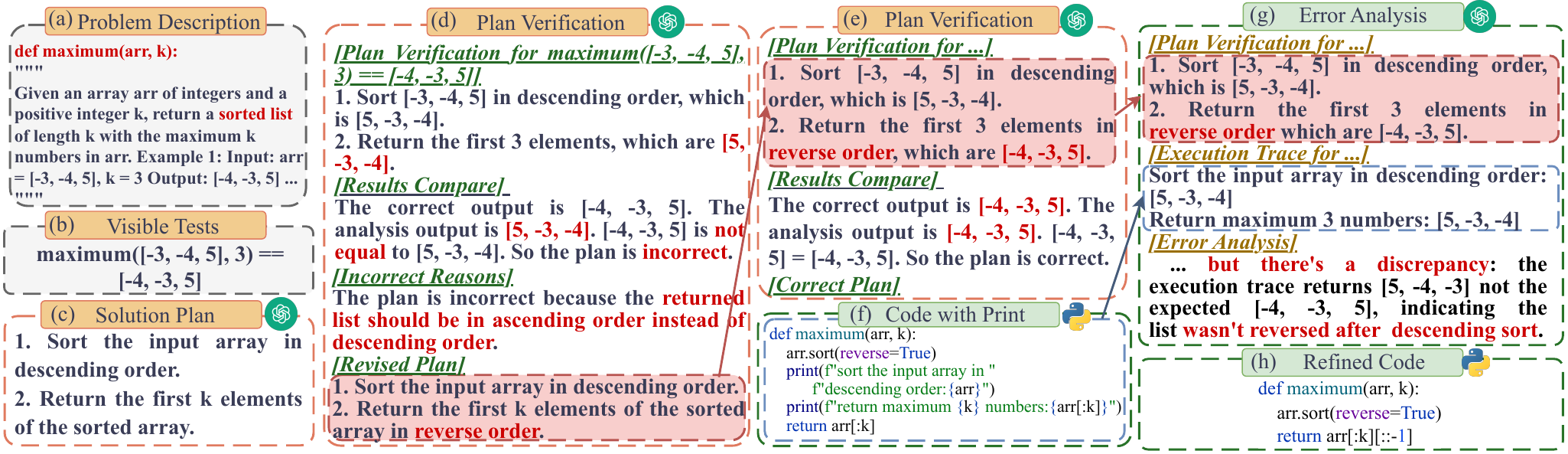}
    \vspace{-0.8cm}
        \caption{A case study of LPW on the \textit{120th} problem in HumanEval, extending from Figures \ref{workflow_solution_generation} and \ref{workflow_code_implemtation}, using GPT-3.5. We omit certain components in Figures \ref{workflow_solution_generation} and \ref{workflow_code_implemtation}, e.g., the plan verification check and the initial code, and present incomplete prompts and responses to save space.}
        \vspace{-0.5cm}
    \label{table:case_study}
    \end{center}
\end{figure*}

\section{Cost-Performance Analysis}
Figure \ref{token_cost_whole} compares Pass@1 accuracy against the average token cost per program for LDB and LPW across five benchmarks using GPT-4o. When analyzing the cost for LDB,  we include the tokens used to generate the seed programs, which account for about 2\% of its total token consumption. LDB consumes fewer tokens per problem but achieves lower accuracy. When measured by the accuracy-cost ratio, computed as Pass@1 accuracy divided by the total tokens used, LDB shows better ratios on HumanEval and MBPP benchmarks. On LiveCode, APPS, and CodeContests benchmarks, LDB and LPW exhibit similar token usage per problem, while LPW  displays notably higher accuracy.  As a result, LPW realizes higher accuracy-cost ratios of 0.60\% on LiveCode, 0.43\% on APPS, and 0.18\% on CodeContests per 1000 tokens, compared to LDB, which reports 0.50\% on LiveCode, 0.39\% on APPS and 0.14\% on CodeContests per 1000 tokens. The lower ratios of LDB arise from insufficient refinements, where multiple ineffective debugging iterations consume significant token resources, yet the generated program remains flawed.

Figure \ref{token_performance} illustrates the variation in Pass@1 accuracy with token consumption across different approaches on the APPS and CodeContests benchmarks using GPT-4o. We introduce Repeated Sampling as an enhanced Baseline. For each problem, it repeatedly samples program solutions from the LLM until either the token consumption exceeds that of LPW, or a solution passes all visible tests and is validated on hidden tests. Repeated Sampling and LDB initially improve accuracy with increased tokens but show negligible improvement after around $10^{3.8}$ tokens on the APPS benchmark and $10^{3.3}$ tokens on the CodeContests benchmark. In contrast, LPW starts with high token consumption for initial plan and verification generation, resulting in a sharp accuracy increase that quickly surpasses Repeated Sampling and LDB after around $10^{4.6}$ tokens on the APPS benchmark and $10^{4.8}$ tokens on the CodeContests benchmark. Repeated Sampling is allocated the same token budget as LPW, while its final accuracy remains lower than LPW on both benchmarks, highlighting the benefits of plan and plan verification in generating high-quality initial code and subsequent refinements. Challenging benchmarks align with LPW usage scenarios, where the precise natural language solution described in the plan and plan verification is essential for logical consistency and understanding non-trivial bugs in the program, particularly when problems involve complex logical reasoning steps.

\section{Case Study} \label{case_study}
Figure \ref{table:case_study} illustrates example message fragments from LPW in the \textit{120th} problem of HumanEval using the GPT-3.5 backbone. LPW successfully  generates the correct program, while all other approaches fail. This problem requires returning a sorted array with the maximum k numbers. However, in the problem description (block (a)), the unspecified order in the output array introduces ambiguity, confusing other methods. LPW struggles at the initial solution plan (block (c)), while the issue is addressed in the \textit{[Revised Plan]}, during plan verification (block (d)). The visible test (block (b)) delineates the reverse order in the return array after sorting in descending order. The initial code with print statements (block (f)) fails on the visible test since the array is not reversed. Subsequently, its execution trace is compared with the plan verification  (block (e)) to identify this bug, as described in the \textit{[Error Analysis]} in block (g). The refined code, which first sorts the array in descending order and then reverses the first k elements into ascending order, successfully solves this problem.

\section{Conclusion}

We introduce LPW, a large language model programming workflow, designed for text-to-code generation tasks. LPW effectively integrates various advanced code generation techniques within a two-phase development model. A key innovation of LPW is the incorporation of plan verification, which enables LLMs to accurately draft an initial program and effectively correct bugs. We evaluate LPW on well-established text-to-code generation benchmarks across various LLMs. LPW significantly improves code generation accuracy compared to other existing approaches and achieves new state-of-the-art Pass@1 accuracy, with 98.2\% on HumanEval, 84.8\% on MBPP, 59.3\% on LiveCode, 62.6\% on APPS, and 34.7\% on CodeContests benchmarks using GPT-4o  as the backbone. In the future, additional visible tests automatically generated by LLMs \citep{chen2022codet} can be explored to improve the performance of LPW.

\section{Limitations}

Similar to other code generation approaches, LPW is constrained by the imperfect reasoning capacity of LLMs. Strengthening the reasoning capacity of LLMs remains an ongoing challenge. While the plan and plan verification have proven valuable across different benchmarks, they require substantial token consumption. In the future, reducing this consumption remains a critical area for improvement. Besides, although the LLM-generated plan and plan verification demonstrate promising accuracy on current tasks, the accuracy of the generated code still lags behind that of the plan and plan verification (Appendix \ref{output_gpt-3.5}). Incorporating alternative solution representations \citep{zelikman2023parsel} alongside natural language representations may assist LLMs in translating text-based solutions into program solutions more accurately.

\bibliography{custom}

\clearpage
\appendix
\section*{Appendix}

\section{Related Work}
\label{Related_Work}

\subsection{Program Synthesis}
Program synthesis remains an open challenge of  generating a program within a target domain-specific language (DSL) from given specifications. One prevalent approach involves searching the large space of possible programs. For example, generalized planning whose solution is formalized as a \textit{planning program} with \textit{pointers} \citep{segovia2024generalized, lei2023novelty}  has demonstrated promising results in synthesizing program solutions for abstract visual reasoning tasks \citep{lei2024generalized} when the DSL is carefully designed. However, hand-crafted DSLs often suffer from limited generalization capacity, and the huge search space diminishes its effectiveness. Recently, large language models trained on vast corpora have excelled in natural language processing (NLP) tasks and have been extended to code generation e.g., GPT-series \citep{achiam2023gpt, openAI2024GPT-4o}, Llama-series \citep{meta2024introducing,roziere2023code,touvron2023llama}, and Claude-series \citep{anthropic2024claude}. LPW leverages the strengths of LLMs in NLP tasks to generate intended solutions in natural language. These text-based solutions demonstrate high-quality logical reasoning steps and satisfactory accuracy, thereby effectively aiding subsequent code generation.

\subsection{Prompting Techniques}
To imitate the logical chain in human brain when tackling reasoning tasks, prompting methods direct LLMs to decompose problems into solvable sub-problems \citep{jiang2023self,zhou2022least,lightman2023let,dhuliawala2023chain} and progressively infer the correct answer with intermediate outputs, as exemplified by chain-of-thought prompting \citep{wei2022chain,kojima2022large}. Inspired by these studies, LPW decomposes a text-to-code problem into several sub-problems described by the solution plan and follows the chain-of-thought prompting idea to verify the solution plan against visible tests with step-by-step analysis. The generated plan and its verification provide step-by-step natural language instructions for code generation, supporting LLMs in both the initial code development and subsequent refinements.  

\begin{figure}
    \begin{center}      
    \includegraphics[width=0.49\textwidth]{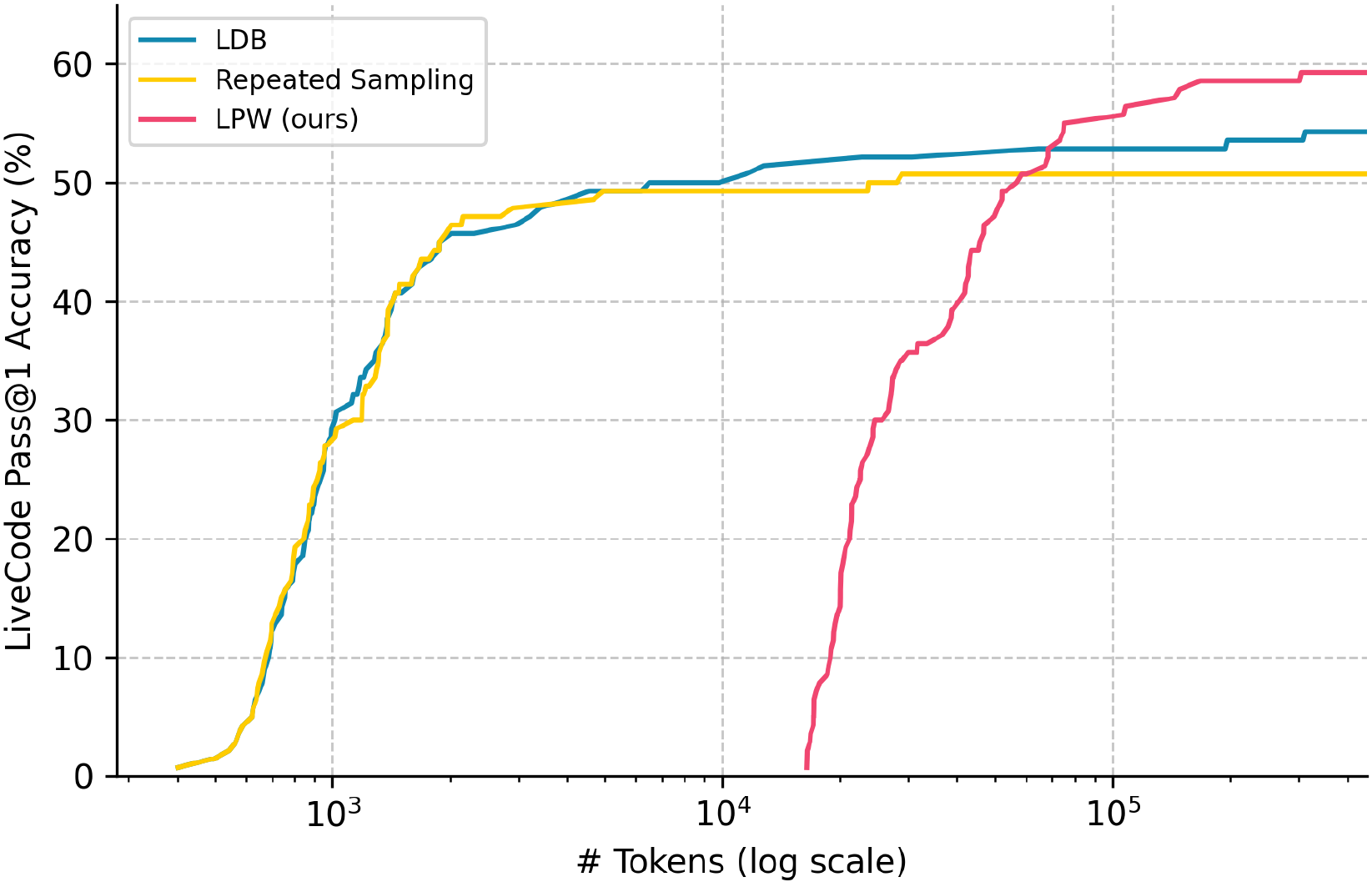}     
    \end{center}
\vspace{-0.3cm}
\caption{Pass@1 accuracy as a function of token consumption for LDB, Repeated Sampling, and LPW on the LiveCode benchmark with GPT-4o.}
    \label{LiveCode_Token}
\end{figure}

\subsection{Code Refinement}  
Accurate program solutions often require iterative refinements due to model limitations \citep{zhong2024ldb, chen2023teaching, shinn2023reflexion}. Various interactive approaches have been proposed to optimize debugging performance in LLMs, such as human feedback \citep{chen2023improving, le2022coderl, wu2024fine}, trained models \citep{huang2023empirical, le2022coderl, yasunaga2021break}, LLM-generated explanations \citep{chen2023teaching, madaan2024self, shinn2023reflexion, tang2023explain}, execution results \citep{zhong2024ldb, holtl2mac, tian2023test}, and  multi-agent collaborations \citep{islam-etal-2024-mapcoder,huang2023agentcoder, qian2023communicative,dong2023self}. Current state-of-the-art LLM debuggers, such as Self-Debugging and LDB, repair various seed programs to create program solutions. However, they encounter difficulties when the initial code substantially deviates from the original intent. Besides, without safeguarding, the refined code frequently diverges from the problem specifications. In contrast, LPW develops initial code that adheres to the validated intended solution through plan verification, minimizing deviations from the problem description. The plan verification further guides the code refinement, ensuring alignment with the problem specifications.

\begin{figure}[t]
    \centering
    \includegraphics[width=1\linewidth]{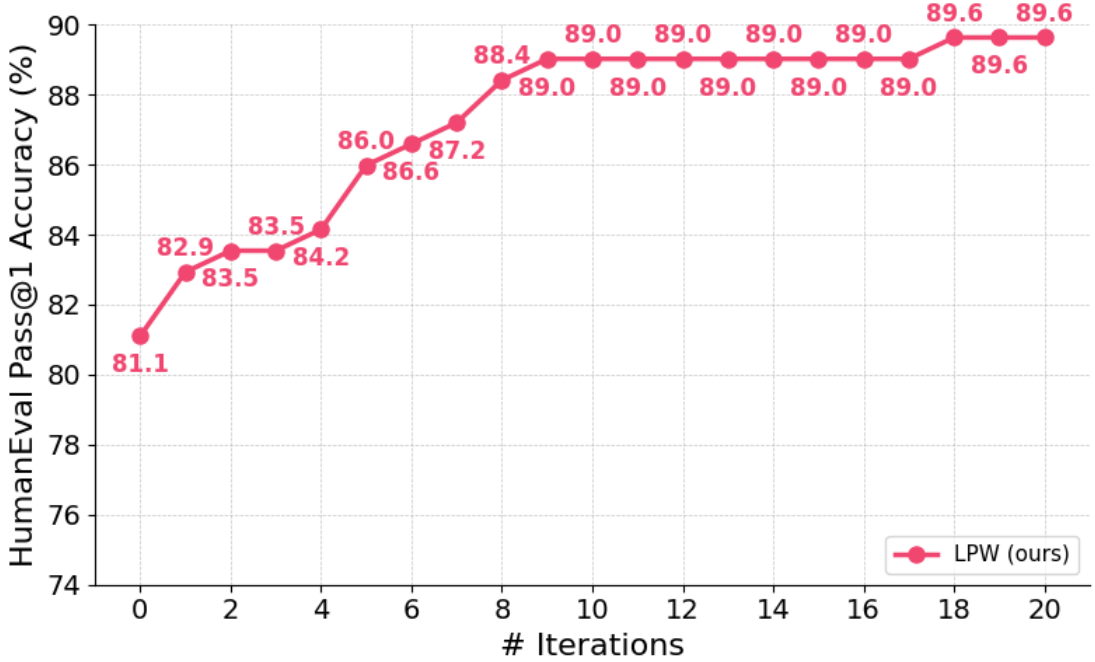}
    \caption{Pass@1 accuracy of LPW on the HumanEval benchmark using GPT-3.5 with 20 iterations in both the solution generation and code implementation phases.
}
    \label{lpw_20_ieration}
\end{figure}

\section{Parameter Study}
\label{Parameter_study}

We conduct an experiment involving 20 iterations for both the solution generation and code implementation phases in LPW. Figure \ref{lpw_20_ieration} shows the variation in Pass@1 accuracy on the HumanEval benchmark using GPT-3.5. The increased number of iterations in the solution generation phase results in higher initial program accuracy, 81.1\%, compared to 79.9\% with 12 iterations before refinements. Subsequently, the accuracy steadily increases, reaching the highest value of 89.6\% after 18 debugging turns, compared to a maximum of 89.0\% with 12 iterations. Overall, a larger number of iterations improves performance in both initial code generation and subsequent refinements. However, the significant token consumption presents challenges for practical applications.

\section{Additional Ablation Study}

\begin{table}[t]
\setlength{\tabcolsep}{5.1pt}
\begin{tabular}{llcccc}
\toprule
       & & \multicolumn{2}{c}{HumanEval}                       & \multicolumn{2}{c}{MBPP}                    \\
\cmidrule(lr){3-4} \cmidrule(lr){5-6}
       & & \multicolumn{1}{c}{Acc}         & \multicolumn{1}{c}{$\Delta$} & \multicolumn{1}{c}{Acc} & \multicolumn{1}{c}{$\Delta$} \\\hline\hline
& LPW    & 89.0 \scriptsize                    & --                    & 76.0              & --                    \\
 \multirow{-2}{*}{GPT-3.5} & LPW-E  &   87.8                     & -1.2                & 75.6          & -0.4                      \\ \hline
\hlineB{2}
\end{tabular}
\caption{Pass@1 accuracy (Acc) for the variant of LPW with the GPT-3.5 backbone. The suffix -E stands for the exclusion of code explanation in the code implementation phase. Other metrics remain consistent with those in Table \ref{table:ablation}.}
\label{table:ablation_eval}
\end{table}

\begin{table*}
\centering
\setlength{\tabcolsep}{7pt}
\begin{tabular}{lcccc}
\hlineB{2}
     & Missing Conditions & Differ from Intended Solution & No Code & Others \\ \hline \hline
LPW  & 33.3                & 5.6                      & 50.0    & 11.1   \\
\hlineB{2}
\end{tabular}
\caption{The percentage of different failure reasons for LPW on the HumanEval benchmark with GPT-3.5 as the backbone. \textit{Missing Conditions} arises from the same issues in the plan and plan verification. \textit{Differ from Intended Solution} indicates the plan and plan verification are manually classified as correct, while the generated code deviates, resulting in failure. \textit{No Code} represents the absence of valid plan and plan verification in the solution generation phase, leading to failure after reaching the maximum number of iterations. \textit{Others} denotes error program solutions caused by various reasons that differ from the previously listed categories.}

\label{unsolved_gpt-3.5}
\end{table*}

\label{Additional_Ablation_Study}

Table \ref{table:ablation_eval} shows the performance of the variant of LPW on the HumanEval and MBPP benchmarks using GPT-3.5 as the LLM backbone. The suffix -E  denotes removing the code explanation when generating the refined program in the code implementation phase. The code explanation facilitates LLMs in aligning text-based error analysis with code implementation when locating and refining incorrect program lines. LPW-E demonstrates a decrease in Pass@1 accuracy on both the HumanEval and MBPP benchmarks.

\begin{figure}[t]
    \centering
    \includegraphics[width=1\linewidth]{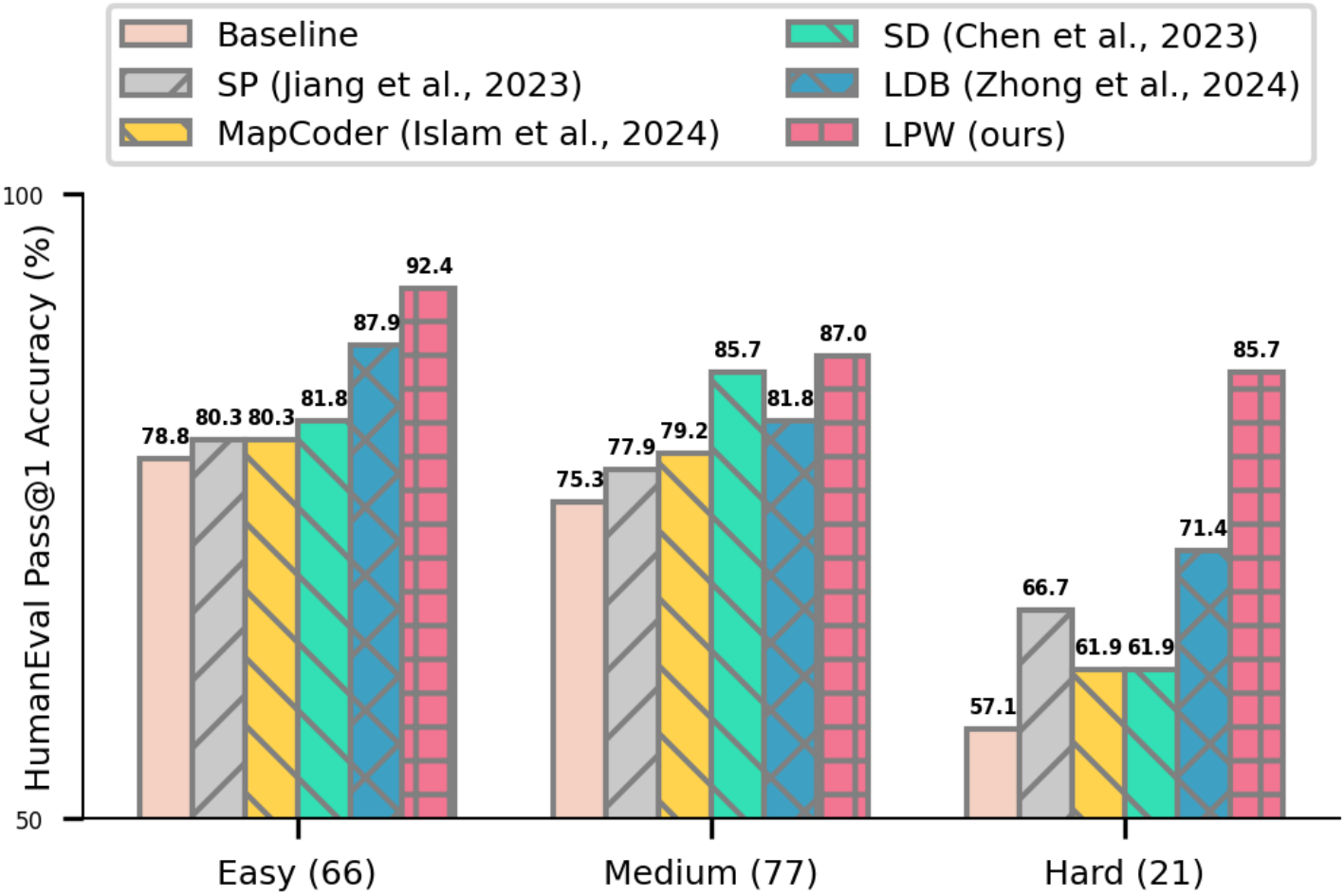}
    \caption{Pass@1 accuracy of Baseline, Self-Planning (SP), MapCoder, Self-Debugging (+Expl) (SD), LDB, and LPW  across different difficulty levels, \textit{Easy}, \textit{Medium}, and \textit{Hard} on the HumanEval benchmark when leveraging GPT-3.5 as the LLM backbone.}
    \label{graph_different_class}
\end{figure}

\section{Analysis of Unsolved Problems for LPW using GPT-3.5}
\label{error_GPT-3.5}

\subsection{Performance Across Different Difficulty Levels}
Figure \ref{graph_different_class} compares the Pass@1 accuracy of competing methods across different difficulty levels, \textit{Easy}, \textit{Medium}, and \textit{Hard}, on the HumanEval benchmark using GPT-3.5. We apply the method described in \citet{zhong2024ldb} to generate the difficulty annotations in Figure \ref{graph_different_class} by querying GPT-3.5 with problem descriptions and canonical solutions. LPW displays convincing performance, exceeding 85\% accuracy across all difficulty levels. For the \textit{Hard} level, LPW achieves 85.7\% accuracy, while competing approaches show a notable degradation, falling below 70\%.

\begin{table}[t]
\centering
\setlength{\tabcolsep}{6pt} 

\begin{tabular}{llcc}
\hlineB{2}
         & & HumanEval   & MBPP \\ \hline \hline
& MapCoder & 18.9 & 46.4  \\
& SD & 22.6  & 36.1  \\
 \multirow{-1.75}{*}{GPT-3.5} & LDB & 28.6 & 37.7 \\
& \cellcolor[HTML]{EEEEEE}\textbf{LPW (ours)} & \cellcolor[HTML]{EEEEEE}\textbf{44.4}  & \cellcolor[HTML]{EEEEEE}\textbf{36.7}\\
\hlineB{2}
\end{tabular}

\caption{The percentage of problems where MapCoder, Self-Debugging (+Expl) (SD), LDB, and LPW generated programs solve the visible tests but fail the hidden tests, out of total failed problems for each method on HumanEval and MBPP, with GPT-3.5 as the backbone.}
        \label{tab:overfit_overfail}
\end{table}

\begin{table}[t]
\centering
        \setlength{\tabcolsep}{6pt} 

\begin{tabular}{llcc}
\hlineB{2}
       &  & HumanEval   & MBPP  \\ \hline \hline
& MapCoder & 4.3 & 13  \\
& SD & 4.3  & 10.4  \\
 \multirow{-1.75}{*}{GPT-3.5} & LDB & 4.9 & 10.4 \\
& \cellcolor[HTML]{EEEEEE}\textbf{LPW (ours)} & \cellcolor[HTML]{EEEEEE}\textbf{4.9} &\cellcolor[HTML]{EEEEEE}\textbf{8.8} \\
\hlineB{2}
\end{tabular}
\caption{The percentage of problems where MapCoder, Self-Debugging (+Expl) (SD), LDB, and LPW generated programs pass the visible tests but fail the hidden tests, out of a total of 164 problems in HumanEval and 500 problems in MBPP, with GPT-3.5 as the backbone.}
        \label{tab:overfit_overall}
\end{table}

\subsection{Failure Reasons}

LPW achieves state-of-the-art performance among the evaluated methods and demonstrates superiority over other LLM debuggers. We categorize the failure reasons for LPW on HumanEval with GPT-3.5 into 4 types. Table \ref{unsolved_gpt-3.5} compares the percentage of different failure reasons out of the total unsolved problems for LPW based on authors' manual review. In LPW, half of the errors result from the \textit{No Code} type, where the generated solution plan fails to be verified on the visible tests, or the resulting verification includes incorrect intermediate outputs in the solution generation phase, leading to failure after reaching the maximum iteration threshold. The second most common reason is \textit{Missing Conditions}, originating from the same issues in the plan and plan verification. For LPW, 5.6\% of failures result from the generated program solution differing from the plan and plan verification (\textit{Differ from Intended Solution}). For example, LPW fails in the \textit{91st} problem, where the generated program is unable to solve the hidden tests due to deviations from the plan and plan verification. The plan verification clearly specifies splitting the input string into sentences using delimiters ``.'', ``?'' or ``!'', but the generated code only handles the full stop case and ignores ``?'' and ``!''.

\begin{figure*}
    \begin{center}
        
    \includegraphics[width=1\textwidth]{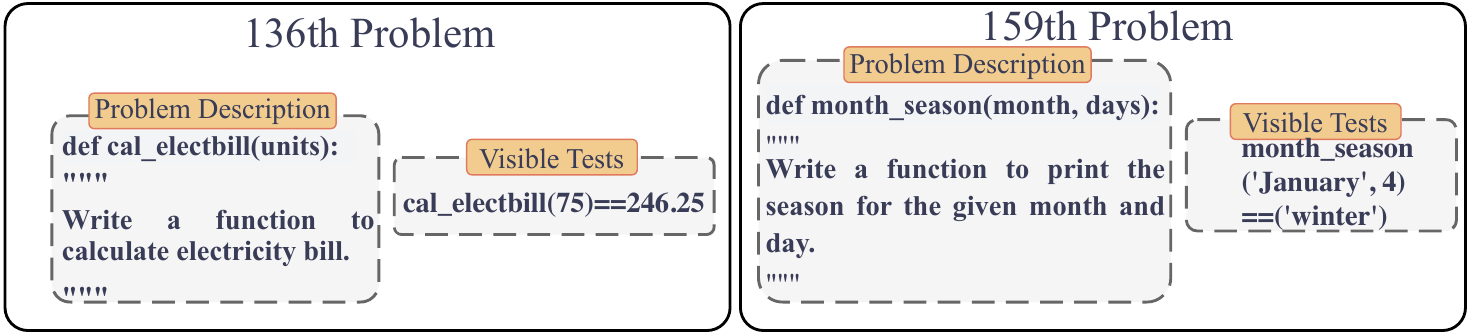}     
    \end{center}

\caption{Example problems in MBPP.}
    \label{mbpp_problme}
\end{figure*}

\subsection{Failure on Hidden Tests}
Tables \ref{tab:overfit_overfail} and \ref{tab:overfit_overall} show the percentage of problems where MapCoder, Self-Debugging (+Expl) (SD), LDB, and LPW generated program solutions pass the visible tests but fail the hidden tests, out of respectively failed problems and the total number of problems in the HumanEval and MBPP benchmarks using GPT-3.5 as the backbone. In Table \ref{tab:overfit_overfail}, 44.4\% of failures in LPW result from solving the visible tests only on the HumanEval benchmark, since except for the \textit{No Code} category, other reasons discussed in Table \ref{unsolved_gpt-3.5} could contribute to this issue. In contrast, less than 30\% of failures in MapCoder, SD and LDB experience this issue on HumanEval as the larger number of failed problems in these methods. In Table \ref{tab:overfit_overfail}, MapCoder demonstrates a higher likelihood of passing only the visible tests, while all other evaluated approaches show similar percentages on the MBPP benchmark, with the remaining failures arising from different reasons. We note that all methods tend to address visible tests only on the same set of problems in the HumanEval benchmark, resulting in the similar percentage out of the total number of problems, as shown in Table \ref{tab:overfit_overall}. In contrast, on the MBPP benchmark, MapCoder exhibits the highest rate of passing only the visible tests out of the total number of problems, consistent with the result in Table \ref{tab:overfit_overfail}, while LPW demonstrates the lowest rate, as shown in Table \ref{tab:overfit_overall}. Meanwhile, all methods are prone to addressing visible tests only on MBPP rather than on HumanEval as indicated in Table \ref{tab:overfit_overall}.   Compared to the detailed problem descriptions in HumanEval, the problem descriptions in MBPP are concise but lack clarity. For example, Figure \ref{mbpp_problme} illustrates two problems in MBPP where LPW generated solutions are tailored to the visible tests but deviate significantly from the canonical solution.

\begin{table*}[t]
\centering
\begin{tabular}{lccc}
\hlineB{2}
     & Plan and Plan Verification & Correct Plan & Correct Plan Verification  \\ \hline \hline
LPW  & 94.5 & 92.7 & 92.7 \\
\hlineB{2}
\end{tabular}
\caption{Percentage of problems where the LLM successfully generates the valid plans and plan verifications in the solution generation phase (first column); percentage of problems where the LLM-generated plans are manually classified as correct (middle column), considering no plan cases; and percentage of problems where the LLM-generated plan verifications are manually classified as correct (last column), considering no plan verification cases. All percentages are reported using GPT-3.5 as the backbone on the HumanEval benchmark, with a total of 164 problems.}
\label{gpt-3.5:performance}
\end{table*}

\begin{table*}[t]
\centering
\begin{tabular}{lcc}
\hlineB{2}
     & Wrong Code $\leftarrow$ Wrong Plan Verification  & Correct Code  $\leftarrow$ Correct Plan Verification \\ \hline \hline
LPW  & 100      & 96.1     \\
\hlineB{2}
\end{tabular}
\caption{The relationship between LLM-generated code solutions and plan verifications on the HumanEval benchmark with GPT-3.5. The first column shows the percentage of problems where the LLM generates incorrect code solutions when plan verifications are incorrect; the second column shows the percentage of problems where correct code solutions are generated when plan verifications are correct. }
\label{gpt-3.5:performance_eval}
\end{table*}

\section{Accuracy of Plans, Plan Verifications, and Programs in LPW using GPT-3.5} \label{output_gpt-3.5}

\subsection{Plans and Plan Verifications}

We manually investigate the accuracy of solution plans and plan verifications generated by GPT-3.5 on the HumanEval benchmark, and the results are presented in Table \ref{gpt-3.5:performance}. Overall, GPT-3.5 generates the correct solution plans and plan verifications in natural language for majority of problems. In LPW, GPT-3.5 successfully produces plans and plan verifications for 94.5\% of the problems. GPT-3.5 generates the correct plans for 92.7\% of the problems and achieves the same accuracy for plan verifications. A common issue in the LLM-generated plan is the omission of certain conditions. For example, solution plan frequently overlooks uppercase situations and negative numbers. We note that the LLM-generated plan verification closely adheres to the solution plan. When the plan is accurate, the verification process strictly follows the plan logic, resulting in a correct analysis. Conversely, if the plan contains logical errors or omits edge cases, the verification process replicates these mistakes. Specifically, for LPW, all correct plans lead to accurate plan verifications, and vice versa.

\subsection{Plan Verifications and Programs}
We further manually explore the relationship between plan verification and program solution on the HumanEval benchmark with GPT-3.5. Table \ref{gpt-3.5:performance_eval} evaluates the conditional probabilities between wrong code and wrong plan verification, as well as between correct code and correct plan verification. Typically, in LPW, accurate plan verifications significantly contribute to correct program solutions, whereas incorrect plan verifications inevitably result in errors. LPW generates program solutions based on plans and plan verifications. Therefore, any accurate descriptions or mistakes, including missed conditions, in the plan and plan verification are propagated to the code. When plan verifications are accurate, 96.1\% of program solutions are correct in LPW. The remaining incorrect instances arise from unclear condition statements in plan verification that fail to account for hidden tests, leading to erroneous program solutions.

The results from Tables \ref{gpt-3.5:performance} and \ref{gpt-3.5:performance_eval} highlight the impressive capability of LLMs in tackling text-to-code generation tasks when outputs are represented in natural language. Plan and plan verification generation accuracy is typically higher than code generation accuracy, underscoring the rationale behind LPW, which produces the high-quality program solution by leveraging plan and plan verification. It is worth exploring methods to help LLMs overcome the challenges of translating natural language solutions into programs, given the strict lexical, grammatical, and semantic constraints. Integrating alternative solution representations \citep{zelikman2023parsel} alongside natural language representations could offer a promising approach.

\section{Refinement Consistency in LPW}
\label{Consistency}

\begin{figure*}
    \begin{center}      
    \includegraphics[width=1\textwidth]{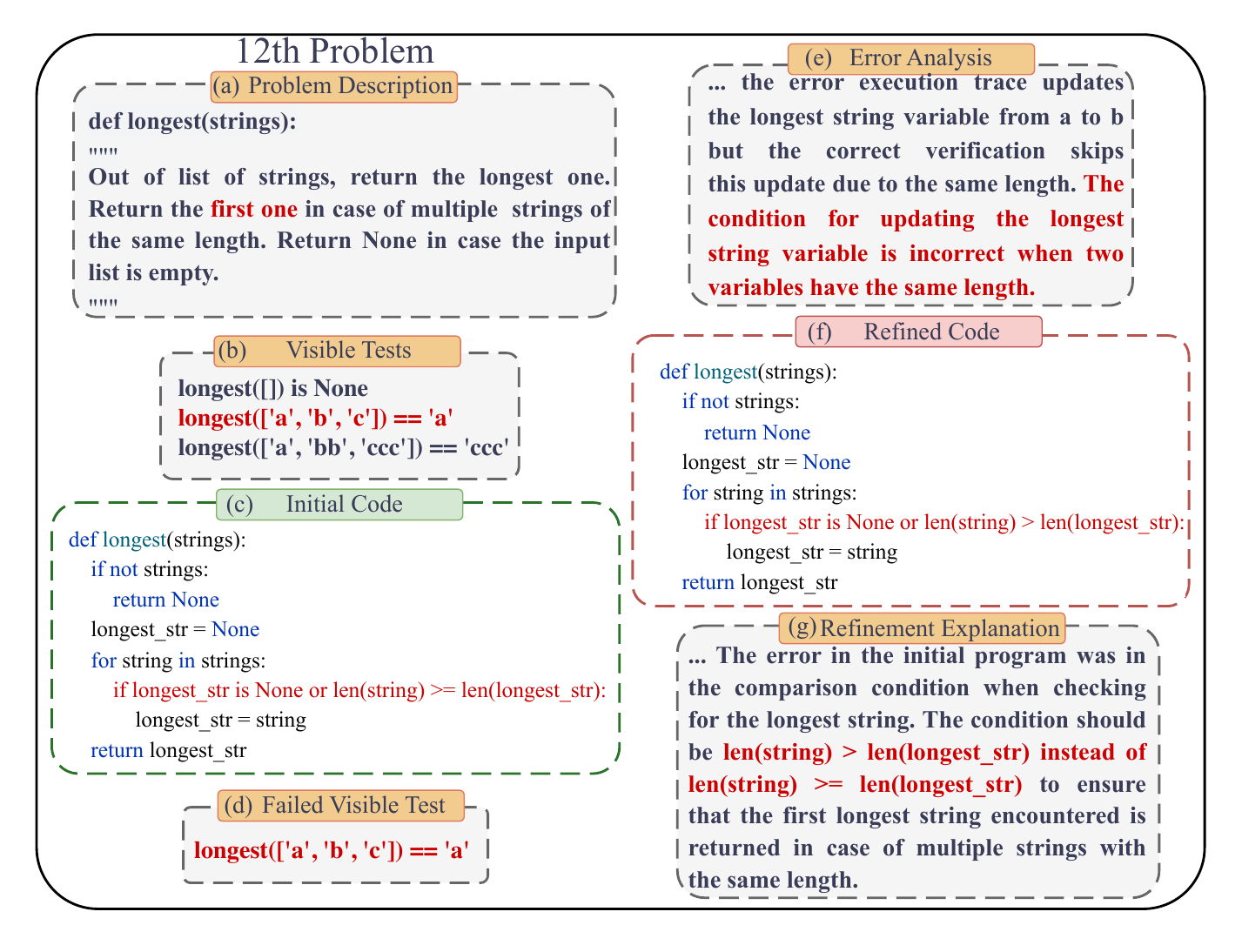}     
    \end{center}

\caption{The \textit{12th} problem in HumanEval, where LPW with GPT-3.5 generated initial code (part (c)) is unable to solve one of the visible tests (part (d)). The refined code (part (f)) successfully solves both visible and hidden tests based on the error analysis (part (e)). The modification in the refined code aligns with the error analysis, as evidenced by the refinement explanation (part (g)).}
    \label{debugging_analysis}
\end{figure*}

LPW allows multiple rounds of debugging to refine code based on error analysis, generated by comparing the code execution trace with plan verification on the failed visible test. LPW queries LLMs to generate refined code accompanied by a refinement explanation, detailing the modifications implemented to address the errors identified in the error analysis. 
For instance, Figures \ref{debugging_analysis} and \ref{table:case_study} illustrate two HumanEval problems where LPW successfully generates the correct program through refinements informed by the error analysis using the GPT-3.5 backbone. We note that in LPW, if the refined code is irrelevant to the error analysis, the entire debugging process degrades to a simple sampling approach, contradicting our original intent. As a result, we manually evaluate the debugging consistency among the generated error analysis (part (e)), the refined code (part (f)), and the refinement explanation (part (g)), as exampled in Figure \ref{debugging_analysis}. LPW demonstrates excellent consistency between error analysis and the refined code, where only one refined code deviates from the error analysis yet still produces the correct solution, across all problems solved through debugging. This result validates the effectiveness of the debugging steps in the code implementation phase for LPW, where the meaningful error analysis enables LLMs to produce the correct program with precise refinements.

\begin{figure}[t]

    \centering
    \centering
    \includegraphics[width=\linewidth]{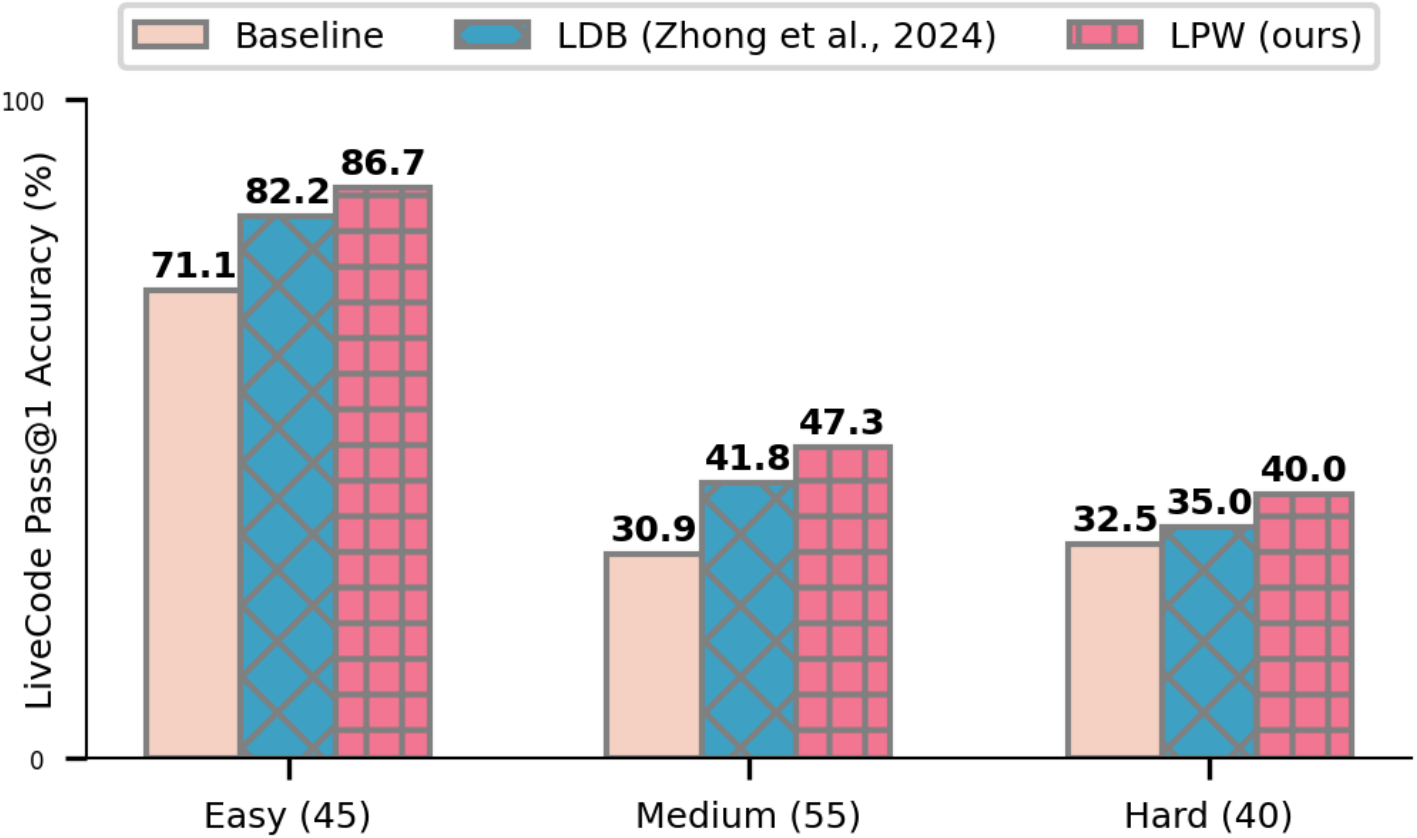}
    \caption{Pass@1 accuracy of Baseline, LDB, and LPW  across different difficulty levels, \textit{Easy}, \textit{Medium}, and \textit{Hard}, on the LiveCode benchmark when using GPT-4o as the LLM backbone.
    }
    \label{LiveCode_Difficulty}
\end{figure}

\begin{table}[t]
    \centering
        \setlength{\tabcolsep}{14pt} 

\begin{tabular}{llc}
\hlineB{2}
& & LiveCode \\ \hline \hline
& LDB & 23.4\\
 \multirow{-2}{*}{GPT-4o} & \cellcolor[HTML]{EEEEEE}\textbf{LPW (ours)} & \cellcolor[HTML]{EEEEEE}\textbf{31.6} \\
\hlineB{2}
\end{tabular}

\caption{The percentage of problems where LDB and LPW generated programs solve the visible tests but fail the hidden tests, out of total failed problems for each method in LiveCode, with GPT-4o as the backbone.}
        \label{tab:4o_errors_unsolved_livecode}
\end{table}

\begin{table}[t]
        \centering
        \setlength{\tabcolsep}{14pt} 

\begin{tabular}{llc}
\hlineB{2}
& & LiveCode \\ \hline \hline
 & LDB & 10.7\\
 \multirow{-2}{*}{GPT-4o} & \cellcolor[HTML]{EEEEEE}\textbf{LPW (ours)} & \cellcolor[HTML]{EEEEEE}\textbf{12.9}\\
\hlineB{2}
\end{tabular}

\caption{The percentage of problems where LDB and LPW generated programs pass the visible tests but fail the hidden tests, out of a total of 140 problems in LiveCode, with GPT-4o as the backbone.}
        \label{tab:4o_errors_all_livecode}
\end{table}

\section{Analysis of Unsolved Problems for LPW using GPT-4o}
\label{error_GPT-4o}
\subsection{HumanEval}

LPW achieves 98.2\% Pass@1 accuracy on HumanEval with the GPT-4o backbone, indicating only 3 unsolvable problems. We further investigate the reasons behind GPT-4o's failures on the \textit{91st}, \textit{132nd}, and \textit{145th} problems as shown in Figures \ref{code_91}, \ref{code_132}, and \ref{code_145}, and attempt to generate the correct program solutions. The \textit{91st} problem fails since GPT-4o misinterprets the linguistic distinction between the word and the letter;  the \textit{132nd} problem's ambiguous description challenges GPT-4o; and the incomplete description of the \textit{145th} problem leads to failed plan verifications. LPW successfully generates correct program solutions for 2 out of 3 problems, achieving 99.4\% Pass@1 accuracy, by involving an additional visible test to validate the intended solution for the \textit{91st} problem and providing a comprehensive problem description for the \textit{145th} problem.

\subsubsection{The \textit{91st} Problem}
\begin{figure*}
    \begin{center}       
    \includegraphics[width=1\textwidth]{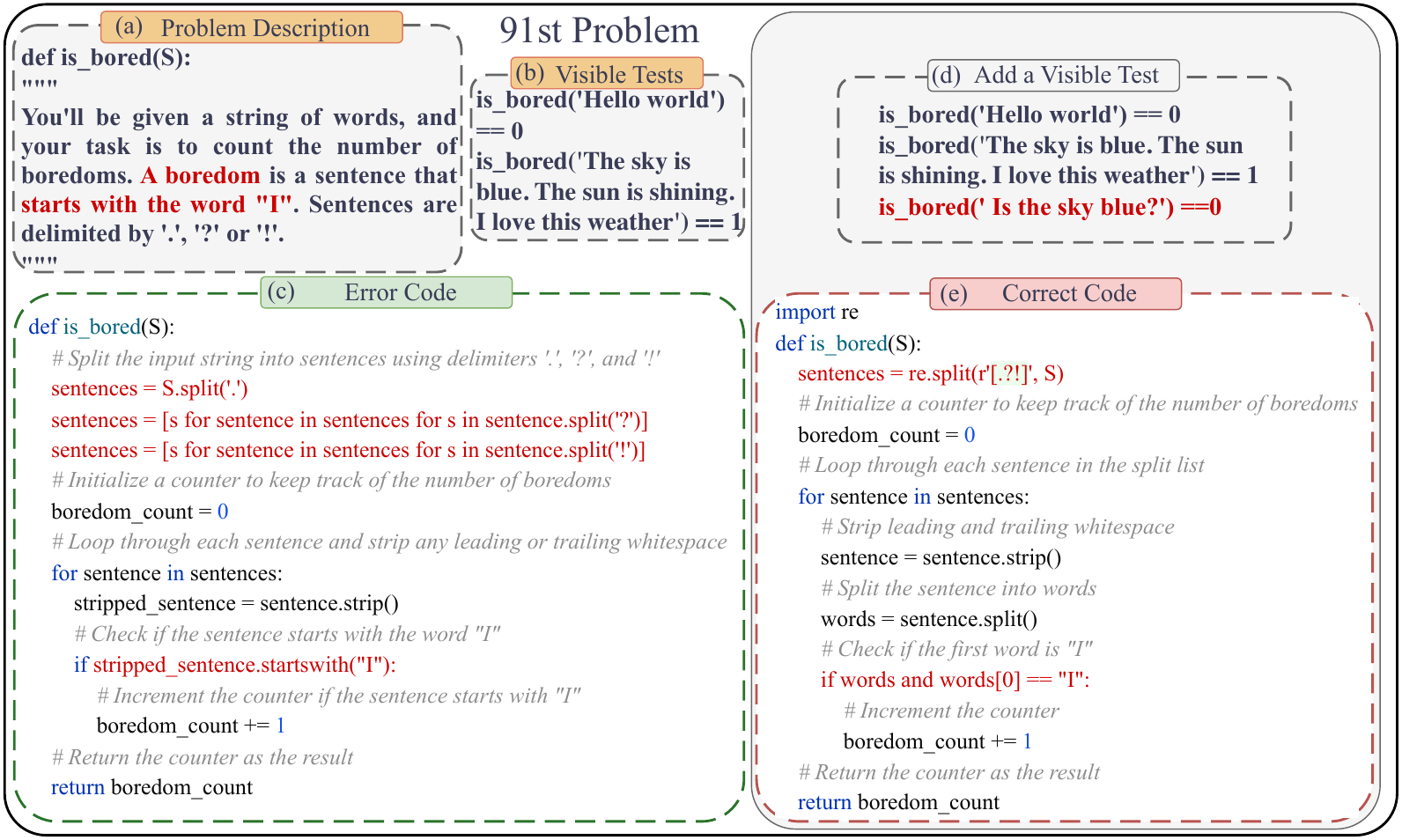}     
    \end{center}

\caption{The problem description (part (a)) and visible tests (part (b)) of the \textit{91st} problem in HumanEval, where GPT-4o generated code (part (c)) addresses the visible tests but fails the hidden tests. However, after converting a failed hidden test to a visible test (part (d)), GPT-4o successfully generates the correct program (part (e)).}
    \label{code_91}
\end{figure*}

Figure \ref{code_91} illustrates the \textit{91st} problem in HumanEval, where the GPT-4o generated code (part (c)) contains an incorrect condition. The code verifies if the sentence starts with the letter ``I'', which is inconsistent with the problem description (part (a)) that requires the sentence to start with the word ``I''. The provided visible tests (part (b)) fail to clarify the correct condition, resulting in the error program  passing the visible tests only.  Inspired by the superior learning-from-test capacity discussed earlier, we convert a failed hidden test into a visible test, highlighted in red in part (d). Consequently, GPT-4o successfully generates the correct program, as shown in part (e). 

\subsubsection{The \textit{145th} Problem}
\begin{figure*}
    \begin{center}
    \includegraphics[width=1\textwidth]{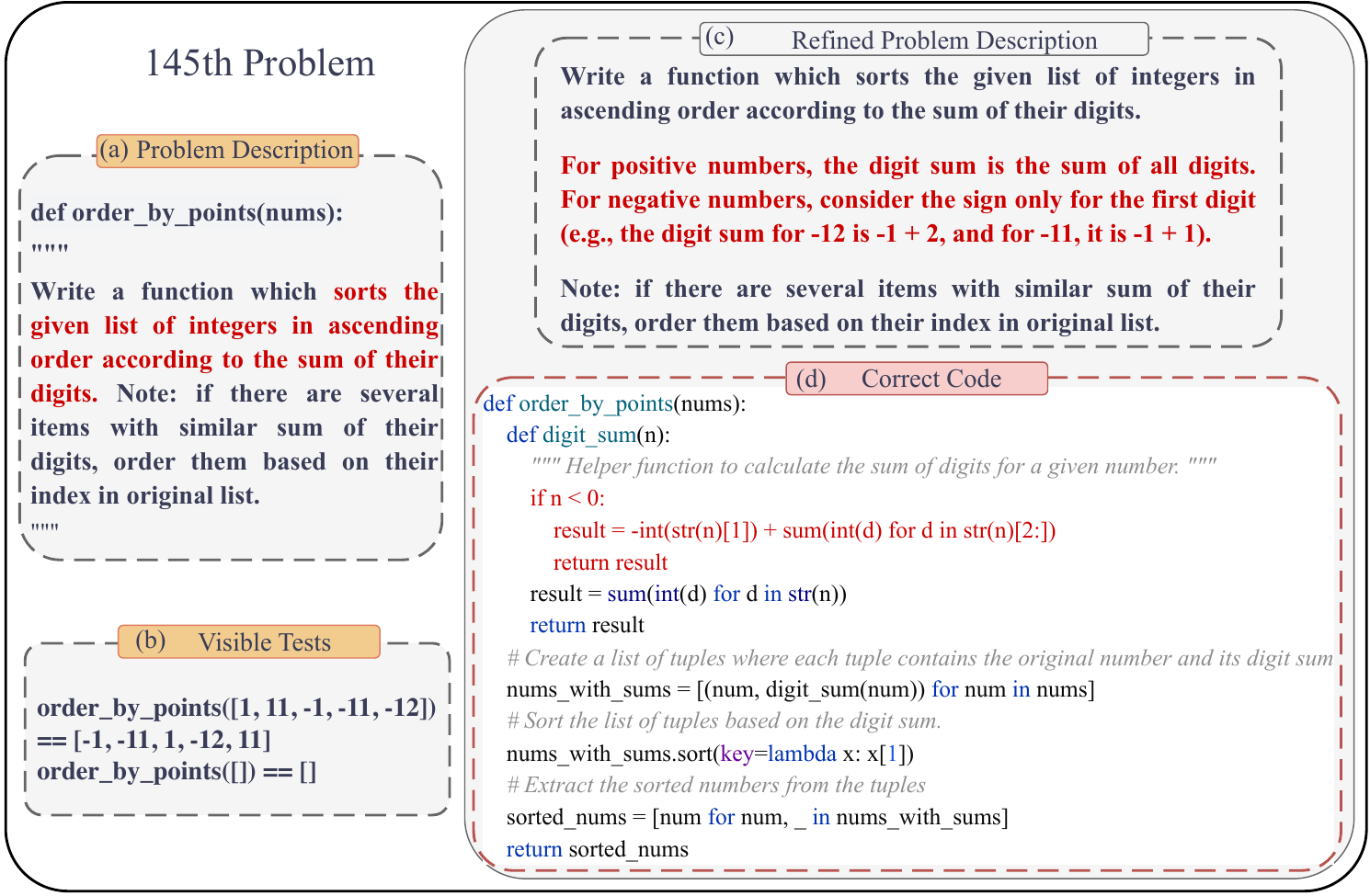} 
    \end{center}

\caption{The problem description (part (a)) and visible tests (part (b)) of the \textit{145th} problem in HumanEval where GPT-4o fails to respond with a valid plan verification, resulting in failure. However, after refining the problem description (part (c)), GPT-4o successfully generates the correct program (part (d)).}
    \label{code_145}
\end{figure*}

Figure \ref{code_145} displays the \textit{145th} problem, where the incomplete problem description (part (a)) results in incorrect plan verification on visible tests (part (b)), leading to a failure after reaching the iteration threshold. The problem description requires returning a list sorted by the sum of digits but omits the specification regarding the sign of negative numbers. This omission confuses GPT-4o, resulting in consistently incorrect sorting when verifying the solution plan on the first visible test. We refine the problem description with a detailed explanation on handling both positive and negative numbers  (part (c)), leading to the correct program solution, as shown in part (d).

\subsubsection{The \textit{132nd} Problem}
\begin{figure*}
    \begin{center}       
    \includegraphics[width=1\textwidth]{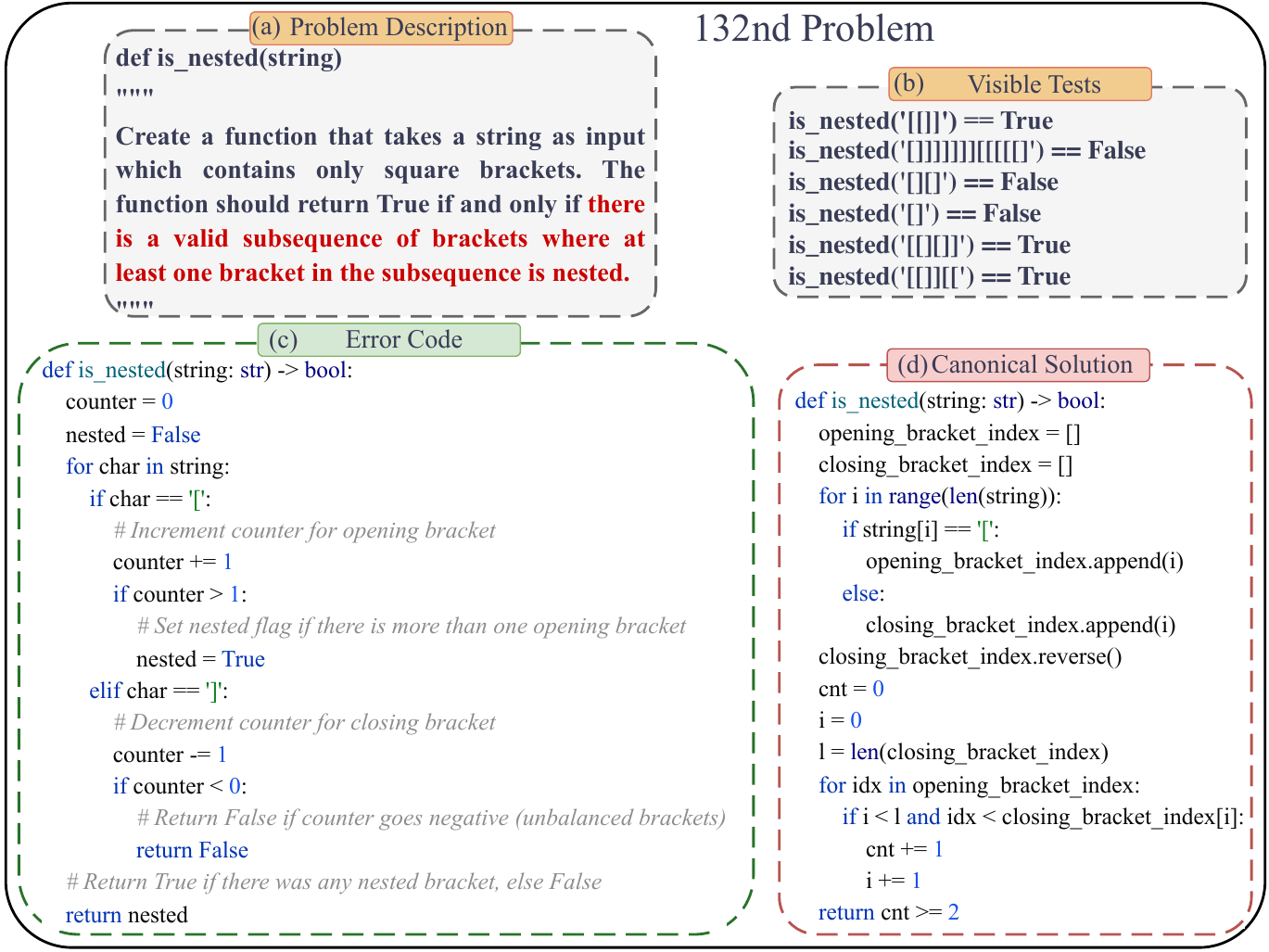}     
    \end{center}

\caption{The problem description (part (a)) and visible tests (part (b)) of the \textit{132nd} problem in HumanEval, where the GPT-4o generated error code (part (c)) passes the visible tests yet fails the hidden tests. GPT-4o consistently generates incorrect programs despite providing additional visible tests or refining the problem description. }
    \label{code_132}
\end{figure*}

Figure \ref{code_132} illustrates the \textit{132nd} problem, where ambiguities in the problem description (part (a)) challenge GPT-4o. The problem description lacks clarity on ``a valid subsequence of brackets'' and fails to specify the meaning of  ``one bracket in the subsequence is nested''. We deduce the intended problem description by prompting GPT-4o with a canonical solution (part (d)). However, the LLM-generated description remains unclear and results in various erroneous programs. Furthermore, adding typically failed hidden tests to the visible tests (part (b))  is also ineffective in clarifying the correct logic. We acknowledge that a clearer description might help generate the correct program. However, some problems are inherently difficult to describe accurately in natural language without careful organization, posing challenges for LLMs.

\begin{table}[t]
    \centering
        \setlength{\tabcolsep}{4.5pt} 

\begin{tabular}{llccc}
\hlineB{2}
         &  & APPS  & CodeContests  \\ \hline \hline
& LDB  & 23.1  & 27.4 \\
 \multirow{-2}{*}{GPT-4o} & \cellcolor[HTML]{EEEEEE}\textbf{LPW (ours)}   & \cellcolor[HTML]{EEEEEE}\textbf{23.1} & \cellcolor[HTML]{EEEEEE}\textbf{29.6} \\
\hlineB{2}
\end{tabular}

\caption{The percentage of problems where LDB and LPW generated programs solve the visible tests but fail the hidden tests, out of total failed problems for each method in APPS and CodeContests, with GPT-4o as the backbone.}
        \label{tab:4o_errors_unsolved}
\end{table}

\begin{table}[t]
        \centering
        \setlength{\tabcolsep}{4.5pt} 

\begin{tabular}{llccc}
\hlineB{2}
      &  & APPS  & CodeContests  \\ \hline \hline
& LDB & 10.8  & 19.3  \\
 \multirow{-2}{*}{GPT-4o} & \cellcolor[HTML]{EEEEEE}\textbf{LPW (ours)}      & \cellcolor[HTML]{EEEEEE}\textbf{8.6} & \cellcolor[HTML]{EEEEEE}\textbf{19.3} \\
\hlineB{2}
\end{tabular}

\caption{The percentage of problems where LDB and LPW generated programs pass the visible tests but fail the hidden tests, out of a total of 139 problems in APPS and 150 problems in CodeContests, with GPT-4o as the backbone.}
        \label{tab:4o_errors_all}
\end{table}

\begin{figure}[t]

    \centering
    \centering
    \includegraphics[width=\linewidth]{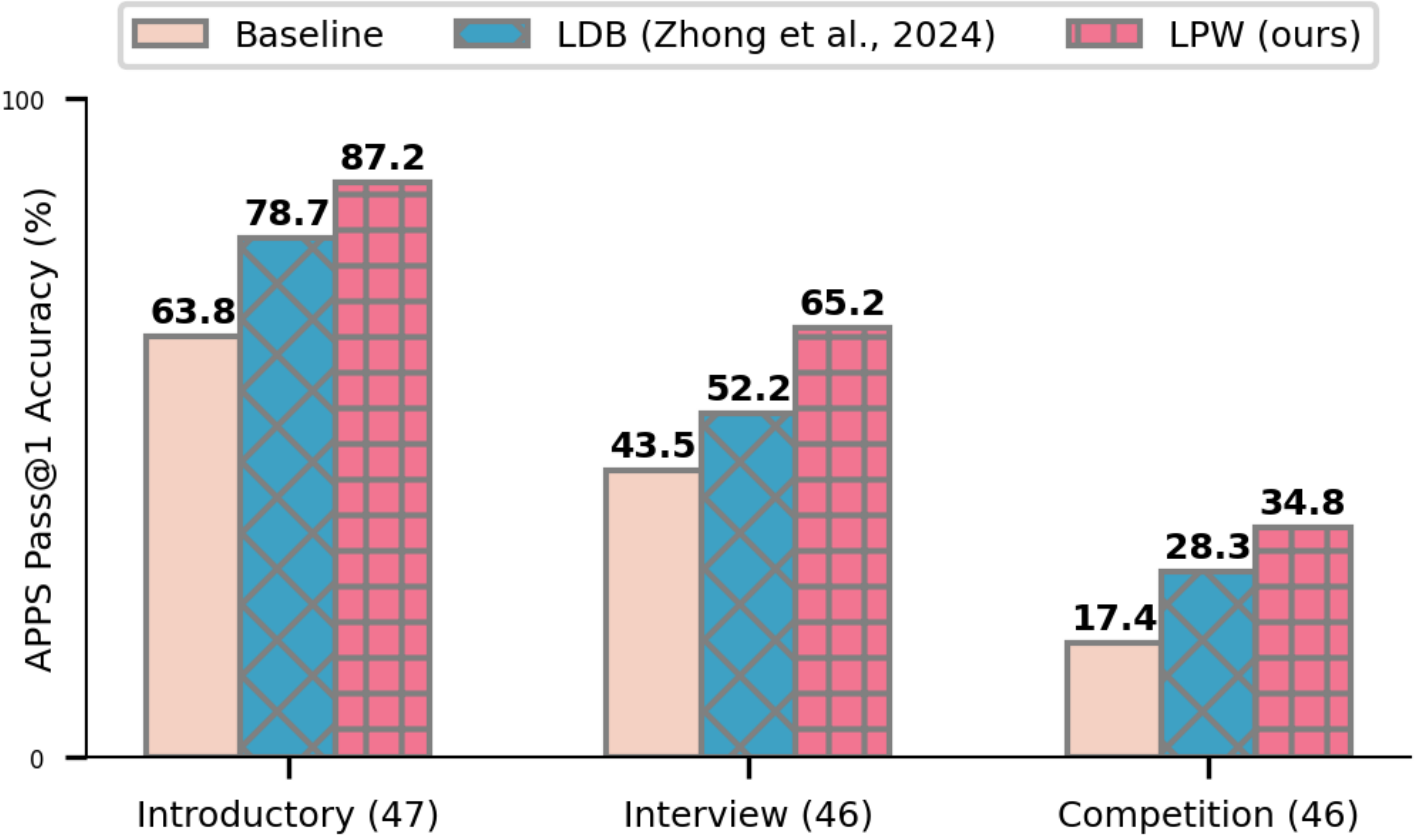}
    \caption{Pass@1 accuracy of Baseline, LDB, and LPW  across different difficulty levels, \textit{Introductory}, \textit{Interview}, and \textit{Competition}, on the APPS benchmark when using GPT-4o as the LLM backbone.  }
    \label{APPS_Difficulty}
\end{figure}

\subsection{LiveCode}

LiveCode, a contamination-free dataset, serves as a reliable benchmark for evaluating code generation performance. It presents challenges for the advanced LLM GPT-4o, with all competing approaches showing performance limitations. We note that LPW consistently surpasses Baseline and LDB by approximately 15\% and 5\%, respectively, in Pass@1 accuracy and across different difficulty levels, as discussed below, emphasizing the reliability of LPW.

\subsubsection{Performance Across Different Difficulty Levels}

Figure \ref{LiveCode_Difficulty} compares the Pass@1 accuracy of Baseline, LDB, and LPW across varying difficulty levels, \textit{Easy}, \textit{Medium}, and \textit{Hard} on the LiveCode benchmark using GPT-4o. LPW achieves the highest Pass@1 accuracy across all levels, surpassing LDB by over 5\% accuracy in each level. Compared to Baseline, LPW delivers over 15\% higher accuracy at the \textit{Easy} and \textit{Medium} levels and 7.5\% higher at the \textit{Hard} level. Conversely, LDB performs similarly to Baseline at the \textit{Hard} level, underscoring its limited refinement capability in addressing more complex tasks.

\subsubsection{Failure on Hidden Tests}

Tables \ref{tab:4o_errors_unsolved_livecode} and \ref{tab:4o_errors_all_livecode} show the percentage of problems where the program solutions generated by LDB and LPW pass only the visible tests but fail the hidden tests out of the total number of failed problems and the total number of problems, respectively, on the LiveCode benchmark using GPT-4o. For LPW, 31.6\% of failures occur when only the visible tests are solved, while for LDB, this percentage is 23.4\%, as shown in Table \ref{tab:4o_errors_unsolved_livecode}. 
In Table \ref{tab:4o_errors_all_livecode}, both LDB and LPW generate a similar proportion of solutions that pass only the visible tests. However, LPW tends to solve only the visible tests in 12.9\% of problems, compared to 10.7\% for LDB.

\subsection{APPS and CodeContests}

\begin{figure*}
    \begin{center}        
    \includegraphics[width=1\textwidth]{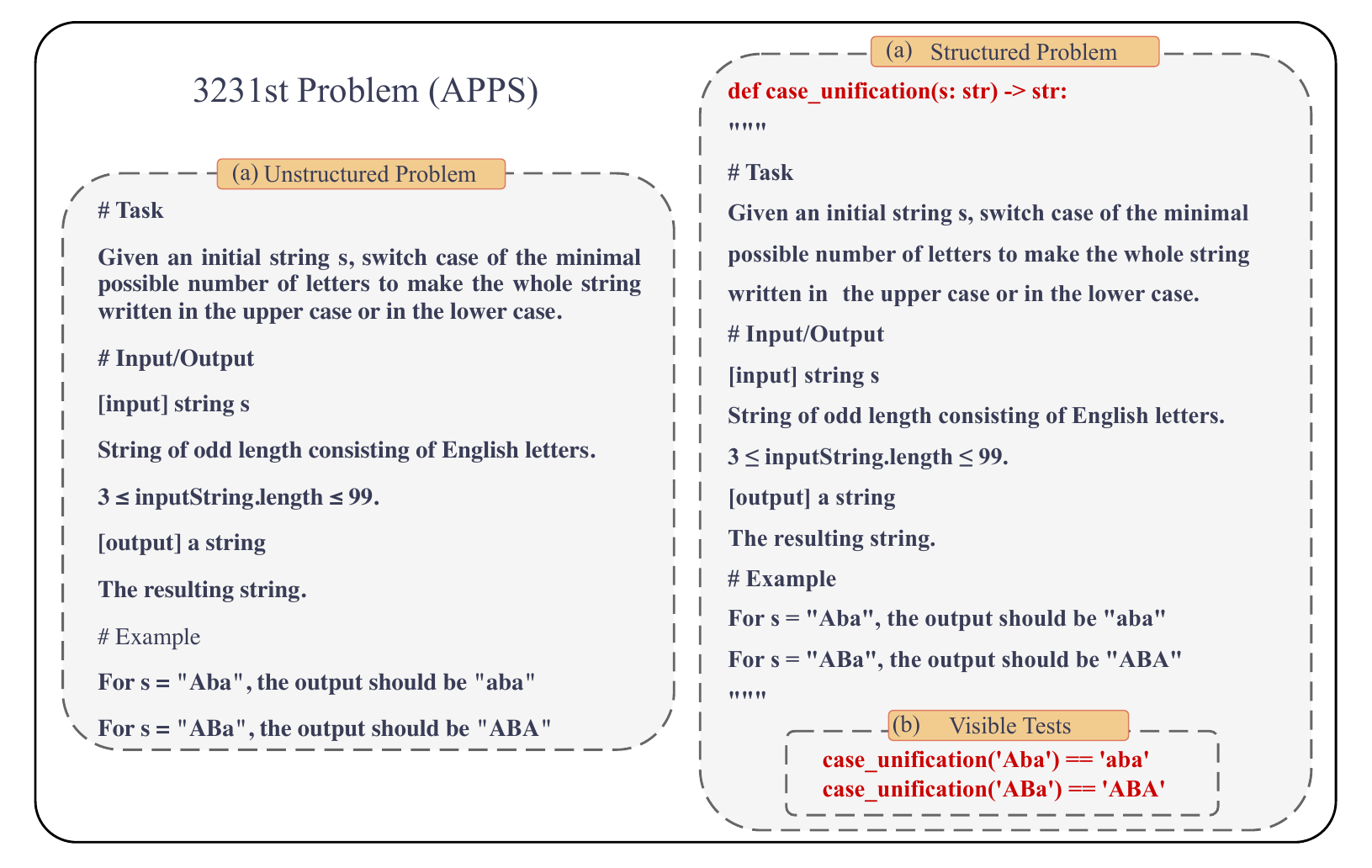}     
    \end{center}

\caption{An example structured APPS problem with a function signature and visible tests, generated by instructing GPT-4o with the unstructured problem description. }
    \label{code_3231}
\end{figure*}

\begin{figure*}
    \begin{center}        
    \includegraphics[width=1\textwidth]{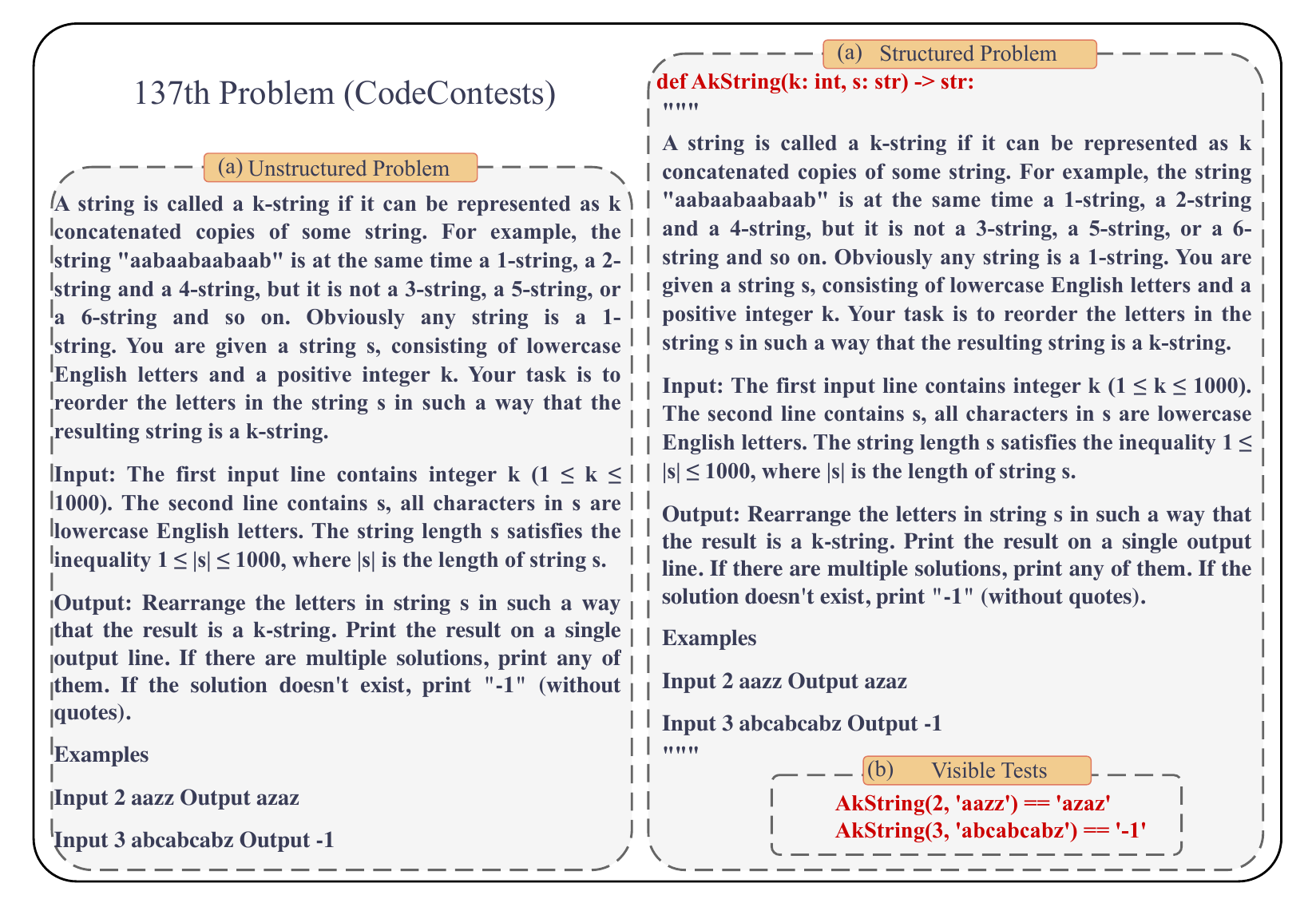}     
    \end{center}

\caption{An example structured CodeContests problem with a function signature and visible tests, generated by instructing GPT-4o with the unstructured problem description. }
    \label{code_137}
\end{figure*}

APPS and CodeContests are unstructured benchmarks where visible tests are intermingled with the problem statements and function signatures are excluded. To align input data structure across benchmarks, we instruct GPT-4o to derive the optimal function signature and identify visible tests for each problem in APPS and CodeContests prior to conducting experiments. Example structured problems from APPS and CodeContests are illustrated in Figures \ref{code_3231} and \ref{code_137}. LPW demonstrates significant improvements on APPS and CodeContests, exceeding around 10\% and 5\% Pass@1 accuracy, respectively, compared to LDB with GPT-4o. However, in contrast to the performance on the HumanEval and MBPP benchmarks, where LPW achieves 98.2\% and 84.8\% Pass@1 accuracy, the 62.6\% accuracy on  APPS and 34.7\% accuracy on CodeContests indicate that even for the advanced LLM GPT-4o, code generation remains challenging when addressing complicated programming problems, such as those encountered in collegiate programming competitions like IOI and ACM \citep{hendrycks2021measuring}.

\subsubsection{Performance Across Different Difficulty Levels}

Figure \ref{APPS_Difficulty} compares the Pass@1 accuracy of Baseline, LDB, and LPW across different difficulty levels, \textit{Introductory}, \textit{Interview}, and \textit{Competition}, on the APPS benchmark using GPT-4o. LPW consistently dominates in Pass@1 accuracy across all difficulty levels. LPW shows strong performance on the relatively easier levels, i.e., \textit{Introductory} and \textit{Interview}, surpassing LDB by around 9\% and 13\% accuracy, respectively, and outperforming Baseline by over 20\% accuracy. For the problems belonging to the most challenging level, \textit{Competition}, LPW achieves 34.8\% accuracy, compared to 28.3\% for LDB and 17.4\% for Baseline. However, all approaches experience a substantial decrease at the \textit{Competition} level, emphasizing the necessity for further improvements. 

\subsubsection{Failure on Hidden Tests}
Tables \ref{tab:4o_errors_unsolved} and \ref{tab:4o_errors_all} present the percentage of problems where the generated program solutions from LDB and LPW solve visible tests but fail hidden tests out of the total failed problems and the total number of problems, respectively, on the APPS and CodeContests benchmarks using GPT-4o as the backbone.  In Table \ref{tab:4o_errors_unsolved}, 23.1\% of failures result from passing only the visible tests on the APPS benchmark, with this percentage increasing to around 30\% on CodeContests for both LDB and LPW. In Table \ref{tab:4o_errors_all}, LDB and LPW display similar percentages of solving visible tests only on each benchmark, ranging from around 10\% on APPS to 19\% on CodeContests. Compared to the results in Table \ref{tab:overfit_overall}, where LDB and LPW address only visible tests in around 5\% of problems on the HumanEval benchmark, LDB and LPW  exhibit weaker performance on the more challenging APPS and CodeContests benchmarks. This is particularly  evident on CodeContests, where the percentage is roughly twice as high as APPS for both LDB and LPW. In APPS and CodeContests, each problem averages approximately 2 visible tests, while CodeContests includes more comprehensive hidden tests, averaging about 23 per problem compared to  only around 5 per problem in APPS, increasing the likelihood of solving only the visible tests.

\subsection{Prompts for LPW}\label{prompts}
We provide the LLM prompts used in LPW in Prompts~\ref{prm:plan_generation} to~\ref{prm:code_refinement}. For conciseness, we only include one example in each prompt.  Full prompts can be found in our released code.

\begin{table*}
\centering
\setlength{\tabcolsep}{3.2pt}
\renewcommand{\arraystretch}{1.2}
\begin{tabular}{lccccc}
\hlineB{2}
     & \multicolumn{2}{c}{Analysis Before Coding} &     \multicolumn{3}{c}{Coding With Debugging}  \\ \cmidrule(lr){2-3} \cmidrule(lr){4-6}
     & Plan                                         & Plan Verification &  Code Explanation & Runtime Information & Intended Solution                                                   \\ \hline\hline
SP   & \textcolor{green}{\ding{51}}  &  \textcolor{red}{\ding{55}}                                      & \textcolor{red}{\ding{55}}    & \textcolor{red}{\ding{55}}
 & \textcolor{red}{\ding{55}}\\
MapCoder   & \textcolor{green}{\ding{51}}  &  \textcolor{red}{\ding{55}}                                      & \textcolor{red}{\ding{55}}    & \textcolor{green}{\ding{51}}
 & \textcolor{red}{\ding{55}}\\
SD   & \textcolor{red}{\ding{55}}  & \textcolor{red}{\ding{55}}                                       & \textcolor{green}{\ding{51}} & \textcolor{red}{\ding{55}}
& \textcolor{red}{\ding{55}}\\
LDB  & \textcolor{red}{\ding{55}}  &  \textcolor{red}{\ding{55}}                                    & \textcolor{green}{\ding{51}} & \textcolor{green}{\ding{51}} &  \textcolor{red}{\ding{55}}       \\
\cellcolor[HTML]{EEEEEE}\textbf{LPW (ours)}  & \cellcolor[HTML]{EEEEEE}\textbf{\textcolor{green}{\ding{51}}}  & \cellcolor[HTML]{EEEEEE}\textbf{\textcolor{green}{\ding{51}}}                                       & \cellcolor[HTML]{EEEEEE}\textbf{\textcolor{green}{\ding{51}}}   & \cellcolor[HTML]{EEEEEE}\textbf{\textcolor{green}{\ding{51}}}
 & \cellcolor[HTML]{EEEEEE}\textbf{\textcolor{green}{\ding{51}}}\\
\hlineB{2}
\end{tabular}
\caption{Features of Self-Planning (SP), MapCoder, Self-Debugging (+Expl) (SD), LDB, and LPW.}
\label{features_stratgies}
\end{table*}

\clearpage
\onecolumn
\lstset{
    basicstyle=\fontsize{10}{10}\selectfont\ttfamily, 
    frame=single,                    
    breaklines=true,                                  
    breakindent=0ex,                                  
    xleftmargin=8pt,                                 
    xrightmargin=8pt,                                
    backgroundcolor=\color[gray]{0.95},               
    keywordstyle=\color{blue}\bfseries,              
    stringstyle=\color{red},                          
    commentstyle=\color{gray},                        
    columns=fullflexible                             
}
\definecolor{mydarkgreen}{RGB}{0, 100, 0}
\definecolor{mydarkred}{RGB}{139, 0, 0}
\lstset{escapeinside={<@}{@>}}
\begin{lstlisting}[
    caption=Prompt for plan generation,
    label=prm:plan_generation,
]
<@\textbf{======================================= System =======================================}@>

You are a Python writing assistant that responds with a step-by-step thought process (IN ENGLISH) to solve Python coding problems. 

<@\textbf{======================================== User ========================================}@>

You will be provided with a series of examples, where each example begins with [Start Example] and ends with [End Example]. In each example, you will be presented with a Python coding problem, starting with [Example Problem Description], which includes the function signature and its accompanying docstring. You will then provide a reasonable solution plan, starting with [Example Start Plan] and ending with [Example End Plan], to solve the given problem.

[Start Example]
[Example Problem Description]
def encrypt(s):
    """ 
    Create a function encrypt that takes a string as an argument and returns a string encrypted with the alphabet being rotated. The alphabet should be rotated in a manner such that the letters shift down by two multiplied to two places.
    """

[Example Start Plan]
Create an alphabet, biased by two places multiplied by two.
Loop through the input, find the letter biased by the alphabet.
Return the result.
[Example End Plan]
[End Example]

... Authors' notes: We omit another example for conciseness. The full prompt can be found in our released code. ...

Lastly, you will be given a Python writing problem, beginning with [Problem Description], which includes the function signature, its docstring, and any potential constraints. The phrase "Let's think step by step" will signal the start of the plan. Your task is to create a solution plan, starting with [Start Plan] and ending with [End Plan].
\end{lstlisting}

\begin{lstlisting}[
    caption=Prompt for plan verification,
    label=prm:plan_verification,
]
<@\textbf{======================================= System =======================================}@>

You are a logical reasoner tasked with performing a step-by-step analysis to derive the correct solution to the given problem based on the provided plan.

<@\textbf{======================================== User ========================================}@>
    
You will be presented with several plan verification examples, each starting with [Start Example] and ending with [End Example]. In each example, you will be given a Python writing problem, starting with [Example Problem Description], followed by the solution plan starting with [Example Solution Plan], and its verification process beginning with [Example Plan Verification for X] for a test case X, starting with [Example Test Cases]. During the verification process, intermediate variables that need to be recorded are clearly identified at the outset, starting with [Record Analysis]. Whenever the value of a recorded intermediate variable is updated, the new result is clearly displayed, beginning with [Record]. After the verification, the derived result is compared to the correct test result, starting with [Results Compare]. If the derived result matches the test result, the output will be [Correct Plan]. If the derived result differs from the test result, the output will be [Incorrect Plan], followed by the reasons for the discrepancy, starting with [Incorrect Reasons], and the revised correct solution plan, beginning with [Start Revised Solution Plan] and ending with [End Revised Solution Plan].

[Start Example]
[Example Problem Description]
def prime_number(n: int):
    """
    In range 0 to 100, returns n-th number that is a prime.
    """

[Example Solution Plan]
1. Iterate number through 0 to 100.
2. Check each number, if it's prime.
3. Keep track of the count of prime numbers found.
4. Stop when we find the n-th prime number.
5. Return the nth prime number.
[Example Test Cases]
assert prime_number(3)==5
[Example Plan Verification for assert prime_number(2)==3]
[Record Analysis]
The return value is the nth prime number, so all nth prime numbers need to be clearly recorded!
1. Call the function prime_number(2).
2. According to line 1 in solution plan, Iterate number through 0 to 100.
3. According to line 2 in solution plan, Check if 0 is prime. It's not.
4. Move to next number 1.
5. According to line 2 in solution plan, Check if 1 is prime. It's not.
6. Move to next number 2.
7. According to line 2 in solution plan, Check if 2 is prime. It is a prime.
8. According to line 3 in solution plan, the count of prime numbers is 1.
[Record]: 1st prime number is 2
9. Move to next number 3.
10. According to line 2 in solution plan, Check if 3 is prime. It is a prime. 
11. According to line 3 in solution plan, the count of prime numbers is 2.
[Record]: 2nd prime number is 3
12. According to line 4 in solution plan, Stop when we find the 2nd prime number.
13. According to line 5 in solution plan, Return the 2nd prime number, which is 3.
[Results Compare]
The test correct output is 3. The logic analysis output is 3. 3=3. Thus, the plan is verified to correctly handle all test cases.
[Correct Plan]
[End Example]

... Authors' notes: We omit another example for conciseness. The full prompt can be found in our released code. ...

Finally, you will be given a problem description, beginning with [Problem Description], along with your generated solution plan, starting with [Solution Plan], to solve the [Problem Description], and multiple test cases starting with [Test Cases]. The phrase "Let's verify the plan" will indicate the beginning of the verification process, followed by your verification steps to confirm whether your generated plan can pass all test cases.
    
For each test case, the verification must include [Record Analysis] to track the intermediate variables at the beginning. If any intermediate variable value is updated during the reasoning process, the updated value should be clearly displayed, starting with [Record]. Please include [Results Compare] to assess the derived outcome against the correct test output. If the derived result matches the test result, output [Correct Plan] and proceed to the next test case. If the derived result does not match the test result, output [Incorrect Plan], followed by the reasons for the discrepancy, starting with [Incorrect Reasons]. Finally, provide the revised solution plan, starting with [Start Revised Solution Plan] and ending with [End Revised Solution Plan], to complete the process.
    
\end{lstlisting}

\begin{lstlisting}[
    caption=Prompt for plan verification check,
    label=prm:plan_verification_check,
]
<@\textbf{======================================= System =======================================}@>

You are a logical reasoner. Your goal is to identify any incorrect logic within the logic verification process.

<@\textbf{======================================== User ========================================}@>

You will be given several examples demonstrating how to evaluate a logic verification process. Each example will begin with [Start Example] and end with [End Example]. In each example, you will find the following: 

[Example Problem Description] outlining the Python writing problem; 

[Example Solution Plan] describing the approach to solve the problem; 

[Example Plan Verification for X], applying the solution plan to a specific test case X. In this process, the intermediate variables to be tracked are analyzed at the start, marked by [Record Analysis]. Whenever the value of a recorded intermediate variable is updated, its new value is displayed starting with [Record]. The [Results Compare] section compares the verification derived result with the correct test output;

[Example Verification Check for X], this section evaluates, step by step, whether the logic verification process for test case X is correct.

If the verification is correct, the output will be [Correct Plan Verification], and  please proceed to the next example. If the verification is incorrect, explanation should be provided and [Incorrect Plan Verification] will be the output to conclude the evaluation.

[Start Example]
[Example Problem Description]
def addOne(message: str):
    """
    You are given a large integer represented as an integer array digits, where each digits[i] is the ith digit of the integer. The digits are ordered from most significant to least significant in left-to-right order. The large integer does not contain any leading 0's. Increment the large integer by one and return the resulting array of digits.
    """

[Example Solution Plan]
1. Convert the list of digits into a number.
2. Increment the number by one.
3. Convert the incremented number back into a list of digits and return it.

[Example Plan Verification for assert addOne([1,2,3])==[1,2,4]]
[Record analysis]
The return value is the incremental resulting array of digits, so the incremental resulting array of digits needs to be clearly recorded!

According to line 1 in solution plan, convert [1,2,3] to the number 123.
According to line 2 in solution plan, Increment 123 by one to get 124.
According to line 3 in solution plan, convert 124 back into the list [1,2,4] 
[Record]: incremental resulting array is  [1,2,4]
According to line 3 in solution plan return incremental resulting array [1,2,4].

[Results Compare]
The test correct output is [1,2,4]. The logic analysis output is  [1,2,4]. [1,2,4]=[1,2,4]. So the plan is verified to correctly handle all test cases.

[Correct Plan]

[Example Verification Check for assert ddOne([1,2,3])==[1,2,4]]:
"Convert [1,2,3] to the number 123" is correct!
"Increment 123 by one to get 124" is correct! since 123+1=124
"Convert 124 back into the list [1,2,4]" is correct!
"return incremental resulting array [1,2,4]" is correct!

In [Results Compare] "The test correct output = [1,2,4]" is correct! "The logic analysis output = [1,2,4]" is correct! The results comparison "[1,2,4]=[1,2,4]" is correct!

All analysis steps are correct!

[Correct Plan Verification]

[Example Plan Verification for assert addOne([-1,2])==[-1,1]]
[Record analysis]
The return value is the incremental resulting array of digits, so the incremental resulting array of digits needs to be clearly recorded!
According to line 1 in solution plan, convert [-1,2] to the number 12.
According to line 2 in solution plan, Increment 12 by one to get 13.
According to line 3 in solution plan, convert 13 back into the list [1,3] 
[Record]: incremental resulting array is  [1,3]

According to line 3 in solution plan return incremental resulting array [1,3].

[Results Compare]
The test correct output is [-1,1]. The logic analysis output is [-1,1]. [-1,1]=[-1,1]. So the plan is verified to correctly handle all test cases.
[Correct Plan]

[Example Verification Check for assert addOne([-1,2])==[-1,1]]:

"Convert [-1,2] to the number 12" is incorrect. The analysis doesn't correctly interpret the -1 and assumes all values are positive, the sequence -1, 2 should form -12.
"Increment 12 by one to get 13" is correct, but as established, the initial conversion should not yield 12. 
"Convert 13 back into the list [1,3]" is correct!
"Return incremental resulting array [1,3]" is correct!

In [Results Compare] "The test correct output = [-1,1]" is correct! "The logic analysis output = [-1,1]" is incorrect! The logic analysis result is [1,3] mentioned in the verification "return incremental resulting array [1,3]". The results comparsion "[-1,1]=[-1,1]" is incorrect! The logic analysis result is [1,3] and [-1,1] is not equal [1,3].

The logic verification process for addOne([-1,2])==[-1,1] is incorrect. The analysis doesn't correctly interpret the -1 and assumes all values are positive, the sequence -1, 2 should form -12. The logic analysis output = [-1,1] is incorrect! It is [1,3]. The results comparison is incorrect since [-1,1] is not equal [1,3].

[Incorrect Plan Verification]

[End Example]

... Authors' notes: We omit another example for conciseness. The full prompt can be found in our released code. ...

Finally, you will be given a problem description, beginning with [Problem Description], followed by your generated solution plan, starting with [Solution Plan], to address the [Problem Description]. You will then work through multiple Plan Verification, each starting with [Plan Verification for X], where X represents a test case. At the start of the verification process, [Record Analysis] examines the intermediate variables that should be tracked. During the logic verification, the tag [Record] indicates any updates to the values of the recorded intermediate variables. The [Results Compare] section documents the comparison between the  verification derived result and the expected test output.

The phrase "Let's evaluate the verification" will indicate the start of the evaluation for each verification process. This will be followed by your step-by-step verification check to assess whether each intermediate output in the verification process is correct, starting with [Verification Check for X], as shown in the examples. If all intermediate results in the verification process are correct, the output will be [Correct Plan Verification], and you will proceed to the next verification. If the verification process is incorrect, an explanation should be provided, and [Incorrect Plan Verification] will be output to conclude the evaluation.

\end{lstlisting}

\begin{lstlisting}[
    caption=Prompt for initial code generation,
    label=prm:initial_code,
]
<@\textbf{======================================= System =======================================}@>

You are a Python writing assistant that only responds with Python programs to solve a Python writing problem.

<@\textbf{======================================== User ========================================}@>

You will receive several examples, each structured as follows, beginning with [Start Example] and ending with [End Example]. Within each example, you will find a Python programming problem starting with [Example Problem Description] and a solution plan starting with [Example Solution Plan]. Additionally, you will receive plan verifications for specific test cases. For each test case X, the plan verification is labeled as [Example Plan Verification for X], providing a detailed logical breakdown and variable value updates, which are recorded starting with [Record]. Following the verification, you will encounter the example-generated program starting with [Example Generated Program]. The program, marked from [Start Program] to [End Program], is generated based on the solution plan and plan verification, ensuring that the program's execution aligns with the plan verification when test case X is used as input. 
    
[Start Example]

[Example Problem Description]
from typing import List
def get_closest_transition_character(word):
    """
    You are given a word. Your task is to find the closest transition character from the right side of the word(case sensitive). The transition character is lowercase and the character after it is uppercase. If no such lowercase character is found, return an empty string.
    >>> get_closest_transition_character("eAsy") == "s"
    """

[Example Solution Plan]
1. Reverse iterate through the characters of the word starting from the last character from the right.
2. For each character, check if the current character is lowercase and if the character after it is uppercase.
3. If step 2 is satisfied, 
4. return the current vowel character.
5. If no such vowel is found, return an empty string.

[Example Plan Verification for assert get_closest_transition_character("eAsy")=="s"]

[Record analysis]
The return value is the closest transition character, so the closest transition character should be recorded!

1. Call the function get_closest_vowel("eAsy").
2. According to line 1 in the solution plan, reverse iterate the word, from the last character to the first character, so the last character is "y"
3. According to line 2 in the solution plan, "y" is a lowercase but the character after "y" is "s" and "s" is a lowercase.
4. Move to the next character based on the reverse iterate, so the character is "s".
5. According to line 2 in the solution plan, "s" is a lowercase and the character after 's' is 'A', and 'A' is uppercase.
6. According to line 3 in the solution plan, step 2 is satisfied, 
7. [Record]: the closest transition character 's'
8. According to line 4 in the solution plan, return the current lowercase character 's'


[Example Generated Program]
[Start Program]

from typing import List
def get_closest_transition_character(word):
    """ You are given a word. Your task is to find the closest transition character from the right side of the word(case sensitive). The transition character is lowercase and the character after it is uppercase.
    >>> get_closest_transition_character("eAsy") == "s"
    """
    # reverse iterate the word
    for i in range (len(word)-1,-1,-1):
        current_character=word[i]
        if current_character.islower():
            if i!=0:
                after_character=word[i-1]
                if after_character.isupper():
                    return current_character
    return ""

[End Program]
[End Example]

... Authors' notes: We omit another example for conciseness. The full prompt can be found in our released code. ...

Finally, you will be provided with a Python writing problem, starting with [Problem Description]. A solution plan will follow, beginning with [Solution Plan]. Next, you will receive several plan verifications. For each test case X, the plan verification, starting with [Plan Verification for X] provides detailed logical reasoning steps to solve it.

Once the plan verification is provided, the "Let's generate the program" flag indicates the start of Python program generation. You will then need to generate the Python program solution for the problem. The plan verification serves as a constraint during program generation.  It is essential to ensure that the execution of the generated program remains consistent with [Plan Verification for X] when using test case X as input. Additionally, the generated program should incorporate all conditions noted in [Plan Verification for X] to solve test case X. Please ONLY output the generated Python program, starting with [Start Program] and ending with [End Program].

\end{lstlisting}

\begin{lstlisting}[
    caption=Prompt for print statement,
    label=prm:print_generation,
]
<@\textbf{======================================= System =======================================}@>

You are a Python writing assistant that only responds with Python programs with PRINT statements.

<@\textbf{======================================== User ========================================}@>

You'll be provided with several examples structured as follows, beginning with [Start Example] and ending with [End Example]. In each example, you will be given a sample Python program, starting with [Example Python Program]. You will also receive several plan verifications for specific test cases. For a test case X, its plan verification, starting with [Example Plan Verification for X], includes a worded description of the logic used to solve test case X. During the verification, the intermediate variable that needs to be tracked is clearly identified, starting with [Record Analysis] at the beginning, and any updates to its value are recorded, starting with [Record].
    
Following this, you will be shown a Python program that includes detailed print statements, starting with [Example Python Program with Print Statements]. These print statements illustrate how the values of the intermediate variables (described in the plan verification) are modified during program execution, as well as how other variables in the program change. These examples will guide you on where and how to add print statements in your Python program.

[Start Example]

[Example Python Program]
from typing import List
def get_closest_transition_character(word):
    """ You are given a word. Your task is to find the closest transition character from the right side of the word(case sensitive). The transition character is lowercase and the character after it is uppercase.
    >>> get_closest_transition_character("eAsy") == "s"
    """
    for i in range (len(word)-1,-1,-1):
        current_character=word[i]
        if current_character.islower():
            if i!=0:
                after_character=word[i-1]
                if after_character.isupper():
                    return current_character
    return ""

[Example Plan Verification for assert get_closest_transition_character("eAsy")=="s"]
[Record analysis]
The return value is the closest transition character, so the closest transition character should be recorded!

1. Call the function get_closest_vowel("eAsy").
2. According to line 1 in the solution plan, reverse iterate the word, from the last character to the first character, so the last character is "y"
3. According to line 2 in the solution plan, "y" is a lowercase but the character after "y" is "s" and "s" is a lowercase.
4. Move to the next character based on the reverse iterate, so the character is "s".
5. According to line 2 in the solution plan, "s" is a lowercase and the character after 's' is 'A', and 'A' is uppercase.
6. According to line 3 in the solution plan, step 2 is satisfied, 
7. [Record]: the closest transition character 's'
8. According to line 4 in the solution plan, return the current lowercase character 's'

[Example Python Program with Print Statements]
from typing import List
def get_closest_transition_character(word):
    """ You are given a word. Your task is to find the closest transition character from the right side of the word(case sensitive). The transition character is lowercase and the character after it is uppercase.
    >>> get_closest_transition_character("eAsy") == "s"
    """

    print(f"Reverse iterate the word {word}")
    for i in range (len(word)-1,-1,-1):
        current_character=word[i]
        print(f"current character at index {i} is {word[i]}")
        if current_character.islower():
            print(f"current character {word[i]} is lowercase")
            if i!=0:
                print(f"There is a character after {word[i]}")
                after_character=word[i-1]
                print(f"character after {word[i]} is {word[i-1]}")
                if after_character.isupper():
                    print(f"character is {word[i-1]} is uppercase")
                    print(f"[Record]: the closest transition character {word[i]}")
                    print(f"Return the closest transition character {word[i]}")
                    return current_character

    print(f"no such lowercase character is found, return an empty string")
    return ""
[End Example]

... Authors' notes: We omit another example for conciseness. The full prompt can be found in our released code. ...

Finally, you will be provided with a Python program, starting with [Python Program], along with several plan verifications for specific test cases. For each test case X, the plan verification, starting with [Plan Verification for X], includes a detailed description of the logic used to solve test case X. In the plan verification, the intermediate variables to be tracked are clearly analyzed at the beginning, starting with [Record Analysis], and any updates to these variable values are recorded, starting with [Record].

The phrase "Let's add print statements" signals the start of the process to incorporate print statements into the provided Python program. Your task is to add print statements that track how the variables in the program change. Ensure that the intermediate variable values (as outlined in the plan verification) are printed using these statements. Output your program with print statements, starting with [Start Program] and ending with [End Program].

\end{lstlisting}

\begin{lstlisting}[
    caption=Prompt for code explanation,
    label=prm:code_explanation,
]
<@\textbf{======================================= System =======================================}@>

You are a Python interpreter tasked with providing an explanation for each line of the Python program.

<@\textbf{======================================== User ========================================}@>
    
You will be given a Python program, and your task is to generate a word-by-word explanation describing the effect of each line in the program. You will be provided with several examples, each beginning with [Start Example] and ending with [End Example]. In each example, you will receive a Python programming problem, starting with [Example Problem Description], along with an example Python program, marked as [Example Python Program], which is generated to solve the given problem. Additionally, you will be provided with an explanation of each line in the example Python program, starting with [Example Explanation For Each Line]. 

[Start Example]
[Example Problem Description]
def encode(message):
    """
    Write a function that takes a message, and encodes in such a way that replaces all letters in the message with the letter that appears 1 place after of that letter in the english alphabet and then delete letter is a vowel. 
    The last letter z is execluded in the message. Assume the input only containing letters. 
    """

[Example Python Program]
def encode(message):
    encoded_message = ""
    for char in message:
        next_char = chr(ord(char) + 1)  
        if next_char not in "aeiouAEIOU": 
            encoded_message += next_char  
    return encoded_message

[Example Explanation For Each Line]
The Python function encode is designed to manipulate a given string (referred to as a message) by replacing the current letter with the letter that appears 1 place after it in the English alphabet and then skipping the letter if it is a vowel:

Function Definition (def encode(message):): Defines a function named encode that accepts one parameter, message. This parameter is intended to be a string that will be processed to create an encoded message.

Initialize Encoded Message (encoded_message = ""): Initializes a variable encoded_message as an empty string. This variable will store the encoded version of the message as characters are processed and approved for inclusion.

For Loop (for char in message:): Iterates over each character in the message. Each character is processed individually.

Calculate Next Character (replacechar = chr(ord(char) + 1)): For each character in the message, this line calculates its next character that appears 1 place after it in the English alphabet. It converts the character to its ASCII value with ord(char), increments this value by 1, and then converts it back to a character with chr().

Check if the resulting character is a Vowel (if replacechar in "aeiouAEIOU":): Check if the resulting character (replacechar) after incrementation is a vowel (either uppercase or lowercase is checked here). If it is a vowel, the continue statement is executed.

Add Character to Encoded Message (else: encoded_message += replacechar): If replacechar is not a vowel, it is appended to encoded_message. This builds up the final encoded string with the modified characters.

Return Encoded Message (return encoded_message): After processing all characters in the original message, the function returns the fully encoded string which consists of all non-vowel characters that are the successors of the original characters in the ASCII sequence.
[End Example]

... Authors' notes: We omit another example for conciseness. The full prompt can be found in our released code. ...

Finally, you will be presented with a problem description, starting with [Problem Description], and your generated Python program, starting with [Python Program], which is meant to solve the [Problem Description]. After this, the "Let's generate the explanation" flag will signal the beginning of the explanation phase. Your task is to generate a word-by-word explanation for each line in the Python program, following the format shown in the previous examples. Please skip the explanation for any line that is a print statement. Output your explanation starting with [Start Explanation] and ending with [End Explanation].

\end{lstlisting}

\begin{lstlisting}[
    caption=Prompt for error analysis,
    label=prm:error_analysis,
]
<@\textbf{======================================= System =======================================}@>

You are a logical reasoner. You will be provided with two logical reasoning processes: [Plan Verification] and [Error Execution Trace]. Your task is to identify any errors in the [Error Execution Trace] by comparing it with the [Plan Verification].

<@\textbf{======================================= System =======================================}@>

You will be provided with several examples, each starting with [Start Example] and ending with [End Example]. In each example, you will receive a Python programming problem, starting with [Example Problem Description], along with an example of an incorrect Python program, marked as [Example Error Program], generated for that problem. You will also be provided with a detailed execution trace of the error program on the failed test case X, labeled as [Example Error Execution Trace for X], including the intermediate variable values.

Additionally, you will be provided with an example of the correct logical reasoning process, labeled as [Example Plan Verification for X]. This process outlines the necessary steps to solve test case X accurately, including condition checks and recording intermediate variable updates, starting with [Record]. Next, [Example Discrepancy Analysis] provides a comparison between the Example Plan Verification and the Example Error Execution Trace, highlighting output differences and identifying where the Error Execution Trace deviates from correctness. Finally, [Example Error Analysis] summarizes the errors identified in the [Example Discrepancy Analysis] and proposes solutions to correct them.

[Start Example]
[Example Problem Description]
def is_palindrome(num):
    """ 
    check if a given integer is a palindrome.
    """

[Example Error Program]
def is_palindrome(num):
    num_str = str(abs(num))
    return num_str == num_str[::-1]

[Example Error Execution Trace for assert is_palindrome(-121)==False]
1. Convert the integer -121 to the string "121" 
2. The integer string "121" is equal to the reversed string "121", the result is True
3. Return True 

[Example Plan Verification for assert is_palindrome(-121)==False]
[Record analysis]
The return value is the checking result about a given integer is a palindrome, so the checking result should be clearly recorded!

1. Call the function is_palindrome(-121).
2. change integer to string, it is "-121"
3. check whether the string "-121" is equal to its reversed string "121-", the checking result is False
4. [Record]: checking result = False
5. Return checking result False


[Example Discrepancy Analysis]
In the plan verification, the recorded value is the checking result:

Let's trace the "checking result" value in the plan verification when it is first-time recorded (SKIP INITIALIZATION).

In the plan verification, the value of checking result is first-time recorded in Line 4 after executing lines:
1. Call the function is_palindrome(-121).
2. change to integer to the string, it is "-121"
3. check whether the string "-121" is equal to its reversed string "121-", the checking result is False
4. [Record]: checking result = False

In the  plan verification, the first-time update changes the checking result value to False.

Let's trace the "checking result" value in the Error Execution Trace.
In Error Execution Trace, the value of checking result is first-time recorded in Line 2 after executing lines
1. Convert the integer -121 to the string "121" 
2. The integer string "121" is equal to the reversed string "121", the result is True

In Error Execution Trace, the first-time update changes the checking result value to True.

The checking result value in the  plan verification and Error Execution Trace are NOT the same, due to False NOT equaling True when the checking result value is first updated. 

Let's carefully analyse the reason with step-by-step thinking:
In lines 1-4 in the  plan verification, the integer -121 is first converted to the string "-121". Then "-121" is compared with its reversed string "121-". "-121" is NOT equaling "121-" so the result is False

In lines 1-2 in Error Execution Trace, the integer -121 is first converted to the string "121". This is different from the plan verification where converting -121 to string is "-121" rather than "121". Then "121" is compared with its reversed string "121". "121" is equaling "121" so the result is True.

[Example Error Analysis]
The error execution trace incorrectly converts the negative integer to its negative integer string. The negative signal is missed. For example, negative integer -121 should be converted to string "-121" but not "121. To fix this error, the negative number must be considered and its negative sign should be contained when converted to string. Such as negative integer -121 should be converted to string "-121".

[End Example]

... Authors' notes: We omit another example for conciseness. The full prompt can be found in our released code. ...

Finally, you will be presented with a problem description, starting with [Problem Description], along with your generated error program, starting with [Error Program], which attempts to solve the [Problem Description]. You will also receive a detailed execution trace, including intermediate variable values, for the failed test case X, starting with [Error Execution Trace for X]. This trace is generated by the error program. Additionally, you will be provided with a correct logical reasoning process, labeled as [Plan Verification for X], which outlines the necessary steps to solve test case X accurately, including condition checks and recording intermediate variable updates, starting with [Record].

Following this, the "Let's do analysis" flag will indicate the start of the analysis phase. Your task is to analyze where the [Error Execution Trace for X] deviates from the [Plan Verification for X], as demonstrated in the examples. This analysis should be output starting with [Discrepancy Analysis].  Finally, you should provide a summary of the errors identified in the [Discrepancy Analysis], including the reasons for these mistakes (IN ENGLISH) and suggestions on how to correct them, starting with [Error Analysis].

\end{lstlisting}

\begin{lstlisting}[
    caption=Prompt for code refinement,
    label=prm:code_refinement,
]
<@\textbf{======================================= System =======================================}@>

You are a Python program fixer. You need to correct an error Python program based on the provided information.

<@\textbf{======================================== User ========================================}@>

You will receive several examples, each structured as follows, starting with [Start Example] and ending with [End Example]. Within each example, you will find a Python programming problem, beginning with [Example Problem Description], followed by an error program provided under [Example Error Program] for the given problem. You will then receive an explanation for the error program, including a line-by-line explanation starting with [Example Error Program Explanation].

Additionally, an error analysis will be provided, starting with [Example Error Analysis], describing the issues in the error program and offering suggestions for refinement. You will then be provided with the refined Python program under [Example Refined Program], based on the error analysis. Following that, a refinement explanation, starting with [Example Refinement Explanation], will be given to show which lines of the program were changed and explain the reasons for those changes.

[Start Example]

[Example Problem Description]
def is_palindrome(num):
    """ 
    check if a given integer is a palindrome.
    """

[Example Error Program]
def is_palindrome(num):
    num_str = str(abs(num))
    return num_str == num_str[::-1]

[Example Error Program Explanation]
Function Definition (def is_palindrome(num):): This line defines a function named is_palindrome that takes one parameter, num. This parameter is expected to be an integer.

Convert Number to Absolute String (num_str = str(abs(num))): A variable num_str is initialized with the absolute value of num converted to a string. The abs() function removes the sign from num if it's negative, ensuring the palindrome check is based solely on the digits.

Check Palindrome and Return (return num_str == num_str[::-1]): This line checks if the string representation of num_str is the same forwards and backwards. It uses the slicing technique [::-1] to reverse the string. If num_str is equal to its reversed version, the function returns True, indicating the number is a palindrome. Otherwise, it returns False.

[Example Error Analysis]
The error execution trace incorrectly converts the negative integer to its negative integer string. The negative signal is missed. For example, negative integer -121 should be converted to string "-121" but not "121. To fix this error, the negative number must be considered and its negative sign should be contained when converted to string.


[Example Refined Program]
def is_palindrome(num):
    num_str = str(num)
    return num_str == num_str[::-1]

[Example Refinement Explanation]
Program line (num_str = str(abs(num))) is changed to (str(num)) to convert the negative integer to its negative integer string by deleting the abs function to keep the negative representation as mentioned in the the error analysis. (str(num)) can correctly convert negative integer -121 to string "-121".

[End Example]

... Authors' notes: We omit another example for conciseness. The full prompt can be found in our released code. ...

You will be presented with a Python writing problem, starting with [Problem Description]. The error program will be provided under [Error Program], followed by an explanation of each line, starting with [Error Program Explanation]. You will then receive an error analysis, starting with [Error Analysis], which describes the issues in the error program and provides refinement suggestions.

The repair process will begin with the phrase "Let's correct the program." Based on the error analysis, generate the refined program. Output your refined program, starting with [Start Refined Program] and ending with [End Refined Program], ensuring that ONLY the Python code is included between these markers. Finally, provide a refinement explanation, starting with [Refinement Explanation], detailing how the program was modified to align with the error analysis.


 
\end{lstlisting}

\end{document}